\documentclass[11pt,a4paper]{article}

\usepackage{subfig}

\usepackage{mathtools}
\usepackage{authblk} % author
\usepackage{algpseudocode,algorithmicx,algorithm}
\usepackage{array}
\usepackage{booktabs}
\usepackage{etoolbox}
\usepackage{tcolorbox}
\usepackage[utf8]{inputenc}
\usepackage[T1]{fontenc}
\usepackage{graphicx}
\usepackage{amsthm}
\usepackage{amsmath}
\usepackage{mathdots}
\usepackage{verbatim}
\usepackage{pifont,url}
\usepackage{enumitem,amssymb}
\usepackage{multirow}
\usepackage{appendix}
\usepackage{epstopdf}
\usepackage{booktabs}
\usepackage{makecell}
\usepackage{diagbox}
\usepackage{setspace}
\usepackage{indentfirst}
\usepackage{qcircuit}

\usepackage{mathrsfs}
\usepackage{latexsym,bm}
\usepackage{amsmath,amsfonts,amsmath,amssymb,amsthm,bbm}
\usepackage{extarrows}

\usepackage{enumerate,cases,multirow}

\usepackage{longtable,colortbl,arydshln,threeparttable}
\definecolor{mygray}{gray}{.9}

\usepackage{indentfirst}
\usepackage{geometry}
\usepackage{cite}

\usepackage{listings}

\usepackage{makeidx}        % generate an index, automatically
\usepackage{booktabs}
\usepackage[unicode=true,bookmarksnumbered,colorlinks,citecolor=red,linkcolor=red,hyperindex,linktocpage=true]{hyperref}
\newcommand{\ket}[1]{| #1 \rangle} % |u>
\newcommand{\bra}[1]{\langle #1 |} % <u|

\def \d {\mathrm{d}}

\def \diag {\text{diag}}

\newcounter{parentalgorithm}

\makeatother

\newtheorem{theorem}{Theorem}[section]
\newtheorem{lemma}{Lemma}[section]

\newtheorem{definition}{Definition}[section]

\theoremstyle{remark}
\newtheorem{remark}{\bf Remark}[section]

\numberwithin{equation}{section}
\usepackage{pifont}

\begin{document}

\title{Quantum Algorithms for Solving Generalized Linear Systems via Momentum Accelerated Gradient and Schr\"odingerization}
\author[1]{Qitong Hu\thanks{Corresponding author: huqitong@sjtu.edu.cn}}
\author[1,3]{Xiaoyang He\thanks{hexiaoyang@sjtu.edu.cn}}
\author[1,2,3]{Shi Jin\thanks{shijin-m@sjtu.edu.cn}}
\author[1,2]{Xiao-Dong Zhang\thanks{xiaodong@sjtu.edu.cn}}
\affil[1]{School of Mathematical Sciences, Shanghai Jiao Tong University, Shanghai, 200240, China}
\affil[2]{Ministry of Education (MOE) Funded Key Lab of Scientific and Engineering Computing, Shanghai Jiao Tong University, Shanghai, 200240, China}
\affil[3]{Institute of Natural Sciences, Shanghai Jiao Tong University, Shanghai, 200240, China}
\date{}
\maketitle
\vspace{-2em}
\begin{abstract}
\par In this paper, we propose a quantum algorithm that combines the momentum accelerated gradient method with Schr\"odingerization [S. Jin, N. Liu and Y. Yu, Phys. Rev. Lett, 133 (2024), 230602][S. Jin, N. Liu and Y. Yu, Phys. Rev. A, 108 (2023), 032603], achieving polynomial speedup over its classical counterpark  in solving linear systems. {The algorithm achieves a query complexity of the same order as the  Schr\"odingerization based damped dynamical system method, namely, linear dependence on the condition number of the matrix, and can overcome the practical limitations of existing non-Schr\"odingerization-based quantum linear system algorithms. These limitations stem from their reliance on techniques such as VTAA and RM, which introduce substantial quantum hardware resource overhead.}
%({\color {green} better to state clearly  what specific significant difficults of the other methods})
Furthermore, it demonstrates both theoretically and experimentally that the auxiliary variables introduced by our method do not dominate the error reduction at any point, thereby preventing a significant increase in the actual evolution time compared to the theoretical prediction. In contrast, the damped method fails to meet this criterion.
%({\color {green} Only see advantage over damped method, not over HHL.})
%, thereby highlighting the superiority of our proposed approach.
%Our analytical framework and conclusions offer
This gives new perspectives for quantum algorithms for linear systems, establishing a novel analytical framework for algorithms with broader applicability, faster convergence rates, and superior solution quality.
\vspace{1.0em}

\par \noindent \textbf{Keywords}: Quantum Simulation, Linear Systems, Momentum Accelerated Gradient Method, Schr\"odingerization Method.
\end{abstract}

\maketitle
\tableofcontents
\section{Introduction}
\par The solution of linear systems $Au=b$ with size $n$ is both a classic and critically important scientific computing problem, with extensive applications across multiple key domains including linear regression analysis, numerical computation of PDEs \cite{Jin2022TimeCA,Jin2025Precondition}, and eigenvalue calculations. However, solving linear systems is often challenged by issues such as matrix ill-conditioningness \cite{Jin2024IllPosed} and the curse of dimensionality \cite{Jin2022TimeCA}. To address these challenges, traditional methods for solving linear systems tend to favor algorithms with low memory consumption, fewer iterations, and strong parallel computing capabilities. For instance: Randomized algorithms (such as randomized Kaczmarz \cite{Strohmer2008Kaczmarz}, stochastic gradient descent \cite{Nemirovski2009SDG,Ghadimi2013SDG}, and Monte Carlo methods \cite{Halton1994MonteCarlo,Wu2019MonteCarlo}) decrease per-iteration computational costs through partial data sampling. Various deep learning-based algorithms \cite{Gu2023DL,Funk2022DL} can perform high-performance parallel computations. However, these requirements undergo significant changes in the quantum computing domain. While mature quantum computers impose fewer constraints on problem scale, they demand quantum-friendly problem formulations - such as requiring matrices to be linear combinations of unitary operators or matrices. This fundamental shift has dramatically altered the direction of algorithm design considerations.
\par In the field of quantum computing, algorithms for linear systems (QLSA) have been extensively studied. Harrow, Hassidim, and Lloyd proposed the HHL algorithm  \cite{Harrow2009QuantumAF}, which utilizes quantum phase estimation and controlled rotation techniques to transform the matrix inversion problem into a quantum state preparation task, theoretically achieving polynomial or even exponential speedup over classical algorithms. The original HHL algorithm has a query complexity of $\mathcal{O}(\kappa^2\delta^{-1})$, where $\kappa$ is the condition number of matrix $A$, and $\delta$ is the target precision. Subsequent research has improved this query complexity.
%through technological innovations.
For example, Childs {\it et al.} \cite{Childs2017SIAM} employed Fourier or Chebyshev fitting based on the Linear Combination of Unitaries (LCU), improving the complexity to a linear dependence on the condition number, i.e., $\mathcal{O}(\kappa\ \text{polylog}(\kappa\delta^{-1}))$. Suba\c{s}{\i} {\it et al.} \cite{Subasi2019PRL} enhanced the adiabatic path design using randomization techniques, achieving near-optimal complexity of $\mathcal{O}(\kappa\log\kappa\delta^{-1})$. Meanwhile, Costa {\it et al.} \cite{Berry2022PRXQuantum} leveraged the discrete adiabatic theorem to theoretically attain the optimal combination of $\kappa$ and $\varepsilon$: $\mathcal{O}(\kappa\log \delta^{-1})$. { However, these algorithms either suffer from slow convergence rates or rely on methods such as VTAA and RM. These techniques involve complex procedures that necessitate a substantial overhead in quantum hardware resources, leading to increased susceptibility to noise and errors on near-term quantum devices\cite{Preskill2018,Jin2025Precondition}.}
%({\color {green} better be more specific abou tthe difficulties.})

\par Recently, another quantum computational approach-Hamiltonian simulation, a field most likely to demonstrate quantum advantage \cite{Feynman1982Quantum}, has garnered increasing attention from researchers \cite{An2023QuantumAF,Jin2024Schrodingerization}. Hamiltonian simulation imposes strict requirements on the Hermitian property of matrices. The Schr\"odingerization framework proposed by Jin {\it et al.} \cite{Jin2023Detailes,Jin2024Schrodingerization} effectively addresses this issue. This method can convert any linear ODE into an equivalent higher-dimensional Schr\"odinger equation, offering a novel perspective for solving a wide range of problems in scientific computing. Numerous algorithms based on the Schr\"odingerization method have been developed, addressing both specific PDEs: such as the fractional Poisson equation  \cite{Jin2025Poisson}, Maxwell's equations  \cite{Jin2023QuantumSO,Ma2024MaxwellTimeDependent}, the Fokker-Planck equation  \cite{Jin2024FokkerPlanck}, and multiscale transport equations  \cite{He2025Multiscale}, as well as broader scientific computing problems, including time-dependent PDEs  \cite{Cao2023QuantumSF} and PDEs with physical boundary conditions  \cite{Jin2024PhysicalBoundary,Jin2025Heat}.
\par The Schr\"odingerization method can also be applied to solve linear systems of algebraic equations \cite{Jin2024QuantumSO}, numerous studies employ the Schr\"odingerization approach for solving linear systems.
%For a linear system arisen from discetization of elliptic PDEs, Hu {\it et al.} \cite{Hu2024QuantumMultiscale} established the fundamental framework: transforming the linear system $Au=b$ into the ODE $\frac{du}{dt}=b-Au$, and provided global error based evolution time estimates under the condition that $A$ is a positive definite matrix. 
Building upon this work, Jin {\it et al.} \cite{Jin2025Precondition} proposed a preconditioning method by introducing another positive definite matrix $B$, thereby reformulating the problem as solving $BAu=Bb$ so the complexity is independent of the condition number of $A$..
Meanwhile, Gu {\it et al.} \cite{Gu2025Helmholtz} extended the differential operator by incorporating a damped wave function, developing the damped dynamical system method, which yielded query complexity estimates similar to those of the HHL algorithm.
Additionally, Hu {\it et al.} \cite{Hu2025preprint} investigated a specific type of linear system formed through iterative scheme and introduced a novel approach for estimating evolution time based on Laplace transform and its inverse transform \cite{Hu2024FundamentalPO,Hu2024TransferPathways,Hu2024Keymotifs}.

\par In this paper, we propose a quantum algorithm based on the momentum accelerated gradient (MAG) and the Schr\"dingerization method. The MAG method \cite{Polyak1964Mom,Lessard2016Mom,Sutskever2013NAG} is an enhanced gradient descent optimization algorithm widely applied to large-scale optimization problems such as deep learning model training. Its core idea involves preemptive updates to the current momentum point during gradient computation, thereby reducing oscillations and achieving faster convergence speeds. Through this method, we obtain query complexity comparable to that of 
%the HHL algorithm
%({\color {green} from Table 1 it seems better than HHL!})
the damped dynamical system method, and it demonstrates advantages over these previous methods in terms of convergence speed or implementation difficulty, {the detailed comparison has been shown in Table \ref{table:1}}.
%({\color {green} what is "practical convergence and stability"?})
{Moreover, our algorithm avoids oscillatory behavior in the solution over time during the time-marching process in most cases (see Fig. \ref{figure:1}), demonstrating superior numerical stability compared to existing Schr\"odingerization-based methods such as the damped method.} This is because, in the design of quantum algorithms, we often need to introduce auxiliary variables to meet the requirements of quantum computation, these variables should not significantly affect the precision of the results. In our method, the introduced auxiliary variables do not dominate the error dynamics, whereas the approach by Gu {\it et al.} \cite{Gu2025Helmholtz} fails to satisfy this property, as will be discussed in details in the following sections.
%{\color {green} I don't see its advantage over HHL etc.  clearly explained.}
\begin{table*}[htbp]
\centerline{
\resizebox{\textwidth}{!}{
\begin{tabular}{p{1.25cm}|p{2.25cm}|p{3.5cm}|p{6.5cm}|p{1.75cm}|p{3.25cm}}
\hline
\hline
    \makecell*[c]{Year}&
    \makecell*[c]{Reference}&
    \makecell*[c]{Query Complexity}&
    \makecell*[c]{Core Idea}&
    \makecell*[c]{Auxiliary\\ Variables}&
    \makecell*[c]{Challenges}\\
\hline
    \makecell*[c]{2009}&
    \makecell*[c]{Harrow {\it et al.}\\  \cite{Harrow2009QuantumAF}}&
    \makecell*[c]{$\mathcal{O}(\kappa^2\delta^{-1})$}&
    \makecell*[c]{First quantum linear system algorithm}&
    \makecell*[c]{-}&
    \multirow{5}{3cm}{\makecell*[c]{\vspace{1em}\\ Rely on highly\\ intricate processes\\ \vspace{1em} \\ Challenge for\\ practical\\ implementation\\  \cite{Jin2025Precondition}}}\\
\cline{1-5}
    \makecell*[c]{2017}&
    \makecell*[c]{Childs {\it et al.}\\  \cite{Childs2017SIAM}}&
    \makecell*[c]{$\mathcal{O}(\kappa\ \text{polylog}(\kappa\delta^{-1}))$}&
    \makecell*[c]{Fourier or Chebyshev fitting based on\\ the linear combination of unitaries}&
    \makecell*[c]{-}&\\
\cline{1-5}
    \makecell*[c]{2019}&
    \makecell*[c]{Suba\c{s}{\i} {\it et al.}\\  \cite{Subasi2019PRL}}&
    \makecell*[c]{$\mathcal{O}(\kappa\log\kappa\ \delta^{-1})$}&
    \makecell*[c]{Randomized adiabatic path method}&
    \makecell*[c]{-}&\\
\cline{1-5}
    \makecell*[c]{2022}&
    \makecell*[c]{Costa {\it et al.}\\  \cite{Berry2022PRXQuantum}}&
    \makecell*[c]{$\mathcal{O}(\kappa\log \delta^{-1})$}&
    \makecell*[c]{Discrete adiabatic theorem}&
    \makecell*[c]{-}&\\
\cline{1-5}
    \makecell*[c]{2020}&
    \makecell*[c]{Shao {\it et al.}\\  \cite{Shao2020Kar}}&
    \makecell*[c]{$\mathcal{O}(\kappa_s^2\log \delta^{-1})$}&
    \makecell*[c]{Row and column iteration method}&
    \makecell*[c]{-}&\\
\iffalse
\cline{1-5}
    \makecell*[c]{2023}&
    \makecell*[c]{Zuo {\it et al.}\\  \cite{Zuo2023Kar}}&
    \makecell*[c]{$\mathcal{O}(\kappa\log \delta^{-1})$}&
    \makecell*[c]{Randomized Kaczmarz method\\ with momentum}&
    \makecell*[c]{-}&\\
\fi
\cline{1-5}
    \makecell*[c]{2024}&
    \makecell*[c]{Li {\it et al.}\\  \cite{Lin2024PRA}}&
    \makecell*[c]{$\mathcal{O}(\kappa_s^2\log \delta^{-1})$}&
    \makecell*[c]{Multirow iteration method}&
    \makecell*[c]{-}&\\
\hline
\hline
    \makecell*[c]{2024}&
    \makecell*[c]{Hu {\it et al.}\\  \cite{Hu2024QuantumMultiscale}}&
    \makecell*[c]{$\tilde{\mathcal{O}}\left( \kappa_g^2\log\kappa_g\log\delta^{-1}\right)$}&
    \makecell*[c]{Gradient descent method}&
    \makecell*[c]{No}&
    \makecell*[c]{Converge slowly}\\
\hline
    \makecell*[c]{2025}&
    \makecell*[c]{Gu {\it et al.}\\  \cite{Gu2025Helmholtz}}&
    \makecell*[c]{$\tilde{\mathcal{O}}\left( \kappa_d\log\kappa_d\log\delta^{-1}\right)$}&
    \makecell*[c]{Damped dynamical system method}&
    \makecell*[c]{Yes}&
    \makecell*[c]{Auxiliary variables\\ affect preceision}\\
\hline
    \makecell*[c]{2025}&
    \makecell*[c]{This paper}&
    \makecell*[c]{$\tilde{\mathcal{O}}\left( \kappa\log\kappa\log\delta^{-1}\right)$}&
    \makecell*[c]{Momentum accelerated gradient}&
    \makecell*[c]{Yes}&
    \makecell*[c]{-}\\
\hline
\hline
\end{tabular}}}
\caption{\label{table:1}\textbf{Comparison of Quantum Algorithms for Solving Linear Systems.} The computational complexity in classical computing is often a polynomial multiple of the matrix size $n$, whereas the query complexity of quantum algorithms in the case of sparse matrices is related to the target precision $\delta$, and the condition number $\kappa$ of matrix $A$ or a similarly defined number. The table lists existing typical QLSAs, among which the latter three are Schr\"dingerization-based. (Non-universal algorithms such as the quantum Jacobi method are not mentioned here \cite{Jin2024QuantumSO}). Here, we ignore the sparsity of the matrix and let $\tilde{\mathcal{O}}$ denote the order ignoring the $\log\log$ term and the discretized parameter $\log N_p$ \cite{Jin2025LinearSystems} introduced by Schr\"odingeration method. Special condition number definitions: $\kappa_s = \|A\|_F \|A^{-1}\|_2$, $\kappa_g = \|A^TA\|_{\max} \|A^{-1}\|_2^2$, $\kappa_d = \|A\|_{\max} \|A^{-1}\|_2$.}
\end{table*}

%{\color {green} Table 1 was never mentioned in the introduction.}

\par The structure of the remaining content of this paper is organized as follows: In Section \ref{section:2}, we introduce the basic framework and termination conditions of the MAG method, as well as how to adapt it into a form suitable for the Schr\"odingerization framework. Meanwhile, in this section, we also provide the fundamental definition of relative convergence and compare it with existing Schr\"odingerization-based linear systems algorithms, highlighting the advantages of MAG. In Section \ref{section:3}, we present the quantum implementation of MAG, including the application of the Schr\"odingerization method and the design of quantum circuits. In Section \ref{section:4}, we conduct simulation experiments under various boundary conditions for the Helmholtz equation and the Biharmonic equation.
\section{Momentum Accelerated Gradient Method}
\label{section:2}
\subsection{Framework for Momentum Accelerated Gradient Method}
\par Let $u_n$ denote the value at the $n$-th iteration step, $u_0$ be the initial iteration value, and $u_\infty=(A^\dagger A)^{-1}A^\dagger b$ be the theoretical solution, which is also the theoretical limit as the iteration tends to infinity. We use the following MAG method \cite{Polyak1964Mom,Lessard2016Mom,Sutskever2013NAG}, with the specific iteration process as follows:
\begin{equation}
    \begin{aligned}
        \label{equ:method:1}
        u_{n+1}&=u_n+\alpha(A^Tb-A^TAu_n)+\beta(u_n-u_{n-1}),
    \end{aligned}
\end{equation}
where $\beta(u_n-u_{n-1})$ is a key step in the MAG method, $\alpha$ is the step size and $\beta$ is the momentum parameter. Under the optimal parameter conditions, it is required that $\alpha=\frac{4}{(\sqrt{L}+\sqrt{\mu})^2}$,  $\beta=\left(\frac{\kappa-1}{\kappa+1}\right)^2$, and $\kappa=\frac{\sqrt{L}}{\sqrt{\mu}}$, where $L=\sigma_{\max}^2$ and $\mu=\sigma_{\min}^2$ are the maximum and minimum eigenvalue of $A^TA$. Note that here, since we assume that the equation $Au = b$ must have a solution, meaning $A$ is invertible, this implies that $A^TA$ is a positive definite matrix, and thus $\mu>0$. In practical applications, it is impossible to estimate the parameters related to $A$ with high precision in specific applications. If one can estimate an upper bound $\hat{L}$ of the same order for $L$, a lower bound $\hat{\mu}$ of the same order for $\mu$, and the corresponding $\hat{\kappa}=\frac{\sqrt{\hat{L}}}{\sqrt{\hat{\mu}}}$, then one can set
\begin{equation}
    \begin{aligned}
        \label{equ:method:2}
        \alpha=\frac{4}{(\sqrt{\hat{L}}+\sqrt{\hat{\mu}})^2},\ \beta=\left(\frac{\hat{\kappa}-1}{\hat{\kappa}+1}\right)^2,
    \end{aligned}
\end{equation}
and achieve the same order of convergence rate as the optimal parameter setting. The detailed proof of this will be provided in the next section.
\par Furthermore, to provide numerical solution of the iterative scheme shown in Eq. (\ref{equ:method:1}), one can set $w_n=[u_n;u_{n-1}]$ and transform it into the following matrix iteration format:
\begin{equation}
    \begin{aligned}
        \label{equ:method:3}
        \tilde{w}_{n+1}=\tilde{\mathbf{H}}\tilde{w}_n+\tilde{\mathbf{F}},
    \end{aligned}
\end{equation}
where the definitions of $\tilde{\mathbf{H}}$ and $\tilde{\mathbf{F}}$ are as follows:
\begin{equation*}
    \begin{aligned}
        \tilde{\mathbf{H}}=\begin{bmatrix}
            (1+\beta)I-\alpha A^TA & -\beta I\\
            I & O
        \end{bmatrix},
        \ 
        \tilde{\mathbf{F}}=\begin{bmatrix}
            \alpha A^Tb\\
            0
        \end{bmatrix}.
    \end{aligned}
\end{equation*}
One can solve this $2n$-dimensional iterative scheme to compute the solution of the linear system $Au = b$.
\begin{remark}
\par The classical Nesterov accelerated gradient method \cite{Nesterov1983Gradient,Nesterov2004Lectures,Sutskever2013NAG} has the following specific form:
\begin{equation*}
    \begin{aligned}
        u_{n+1}&=v_n+\alpha(A^Tb-A^TAv_n),\\
        v_{n+1}&=u_{n+1}+\beta(u_{n+1}-u_n),
    \end{aligned}
\end{equation*}
in which $v_n$ is the extrapolation point, which is also the core of the Nesterov method.
\end{remark}
\subsubsection{Matrix Iteration Format Suitable for Schr\"odingerization}
\par The Schr\"odingerization method is a universal approach; however, if certain restrictions are imposed on the iterative scheme or the ODE to be solved, the range of hyperparameter selection can be broadened. It is generally believed that the Schr\"odingerization method tends to be more stable when all eigenvalues of $\frac{\tilde{\mathbf{H}} + \tilde{\mathbf{H}}^\dagger}{2} - I$ are negative ($\tilde{\mathbf{H}}$ is defined in Eq. (\ref{equ:method:2})), whereas cases with positive eigenvalues have also been studied and addressed \cite{Jin2024IllPosed}. Therefore, here we aim to apply a linear transformation to our obtained iterative scheme to meet this requirement.
\par We define the new variable $w_n = [(1-\beta)u_n; \sqrt{\alpha\beta}Au_{n-1}]$, and thus we can obtain the following iterative scheme:
\begin{equation}
    \begin{aligned}
        \label{equ:method:4}
        w_{n+1}=\mathbf{H}w_n+\mathbf{F},
    \end{aligned}
\end{equation}
in which $\mathbf{H}$ and $\mathbf{F}$ are defined as follows:
\begin{equation*}
    \begin{aligned}
        \mathbf{H}=\begin{bmatrix}
            I-\alpha A^TA & -\sqrt{\alpha\beta} A^T\\
            \sqrt{\alpha\beta} A & \beta I
        \end{bmatrix},
        \ 
        \mathbf{F}=\begin{bmatrix}
            \alpha A^Tb\\
            0
        \end{bmatrix}.
    \end{aligned}
\end{equation*}
To compute the steady-state solution of the iterative company mentioned in Eq. (\ref{equ:method:4}), we can directly calculate $(I-\mathbf{H})^{-1}\mathbf{F}$, thereby obtaining
\begin{equation*}
    \begin{aligned}
        (I-\mathbf{H})^{-1}\mathbf{F}=
        \begin{bmatrix}
            (1-\beta)(A^TA)^{-1}A^Tb\\
            \sqrt{\alpha\beta}A(A^TA)^{-1}A^Tb
        \end{bmatrix}.
    \end{aligned}
\end{equation*}
We can see that the first term is $(1-\beta)$ times the solution we need, while the second term, under the condition that $A$ is an invertible square matrix, equals $\sqrt{\alpha\beta}b$. Since $b$ is known, this can also serve as a validation term for our algorithm.
\par We need to note that the new iterative scheme presented in Eq. (\ref{equ:method:4}) is derived from a similarity transformation for the original iterative scheme in Eq. (\ref{equ:method:4}). Therefore, the stability and convergence rate of the two iterative schemes are identical. Moreover, since $A^TA$ is a positive definite matrix, we have
\begin{equation}
    \begin{aligned}
        \label{equ:method:5}
        \lambda_{\max}\left(\frac{\mathbf{H}+\mathbf{H}^\dagger}{2}-I\right)=\lambda_{\max}\left(\begin{bmatrix}
            -\alpha A^TA & O\\
            O & -(1-\beta)I
        \end{bmatrix}\right)=\max\{-\alpha\mu,-(1-\beta)\}<0.
    \end{aligned}
\end{equation}
The result introduced in Eq. (\ref{equ:method:8}) indicates that the iterative scheme in Eq. (\ref{equ:method:4}) is more suitable for the Schr\"odingerization method.
\subsubsection{Iteration Termination Condition}
\par As the number of iteration steps increases, using Eq. (\ref{equ:method:1}) will continuously approach the exact value $u_\infty$. However, this process does not automatically terminate, so one needs to establish iteration stopping criteria. Jin {\it et al.} \cite{Jin2024QuantumSO} used fidelity to determine the termination step-a result related to the matrix norm. Here, we propose an iteration termination condition based on spectral radius. Let $\Delta w_n=w_n-w_\infty$, through Eq. (\ref{equ:method:4}) one can derive the following error propagation equation:
\begin{equation*}
    \begin{aligned}
        \Delta w_{n+1}=\mathbf{H}\Delta w_n,
    \end{aligned}
\end{equation*}
which demonstrates that $\Delta w_{n+1}=\mathbf{H}^n\Delta w_0$. To further process this error ODE, we first perform a Jordan decomposition of $\mathbf{H}$ as $\mathbf{H}=X^{-1}\Lambda X$, and let the singular decomposition of $A$ be $A=U^\dagger \Sigma V$ with $\Sigma=\diag([\sigma_1,\cdots,\sigma_n])$. Then, the specific expressions for $X$ and $\Lambda$ are as follows:
\begin{equation*}
    \begin{aligned}
        X&=\diag([V,U])
        P\diag([Q_1,Q_2,\cdots,Q_{n-1},Q_n]),\\
        \Lambda&=\diag([\lambda_1^+,\lambda_1^-,\cdots,\lambda_n^+,\lambda_n^-]),
    \end{aligned}
\end{equation*}
in which the permutation matrix $P$ is defined as
\begin{equation*}
    \begin{aligned}
        P_{ij}=\left\{\begin{array}{l}
            1,\text{ $j=\sigma(i)$},\\
            0,\text{ else.}
        \end{array}\right.,\ \sigma(i) = 
        \begin{cases} 
        2i - 1,\text{ $1 \le i \le n$}, \\
        2(i - n),\text{ $n + 1 \le i \le 2n$}.
        \end{cases},
    \end{aligned}
\end{equation*}
the matrix $Q_i$ and $Q_i^{-1}$ of size $2 \times 2$ are
\begin{equation*}
    \begin{aligned}
        Q_i^{-1}=\begin{bmatrix}
            \beta-\lambda_i^+ & \beta-\lambda_i^-\\
            -\sqrt{\alpha\beta}\sigma_i & -\sqrt{\alpha\beta}\sigma_i
        \end{bmatrix},\ 
        Q_i=\frac{1}{\delta\lambda_i}\begin{bmatrix}
            -1 & \frac{-(\beta-\lambda_i^-)}{\sqrt{\alpha\beta}\sigma_i}\\
            1 & \frac{\beta-\lambda_i^+}{\sqrt{\alpha\beta}\sigma_i},
        \end{bmatrix}
    \end{aligned}
\end{equation*}
where $\delta\lambda_i=\lambda_i^+-\lambda_i^-$, and $\lambda_i^+$, $\lambda_i^-$ are
\begin{equation}
    \begin{aligned}
        \label{equ:method:6}
        \lambda_i^{\pm}=\frac{(1+\beta-\alpha\sigma_i^2)\pm\sqrt{(1+\beta-\alpha\sigma_i^2)^2-4\beta}}{2}.
    \end{aligned}
\end{equation}
It can be easily seen that both $\diag([V,U])$ and $T$ are unitary matrices, and one has $\ln\kappa_2(Q_i)\lesssim \ln\hat{\kappa}$. Therefore, we can derive that $\kappa_2(X) \lesssim \ln\hat{\kappa}$, leading to the following error estimate
\begin{equation}
    \begin{aligned}
        \label{equ:method:7}
        \ln\frac{\|\Delta w_n\|_2}{\|\Delta w_0\|_2}\le \ln\|\mathbf{H}^n\|_2\lesssim n\ln\rho(\mathbf{H})+\ln\hat{\kappa},
    \end{aligned}
\end{equation}
where $\rho(\mathbf{H})$ is the spectral radius of $H$. If we define $T$ as the number of iterations $n$ required to satisfy the convergence condition $\frac{\|\Delta w_n\|}{\|\Delta w_0\|}<\delta$ with $\ln\delta=\Theta(\ln\hat{\kappa})$, then the following constraint on $T$ serves as a sufficient condition for meeting this criterion:
\begin{equation}
    \begin{aligned}
        \label{equ:method:8}
        T&\gtrsim \frac{\ln\delta^{-1}}{-\ln\rho(\mathbf{H})}.
    \end{aligned}
\end{equation}
In fact, one can verify that the inequality in Eq. (\ref{equ:method:7}) is tight. We simply let $\Delta w_0 = \xi$, where $\xi$ is the eigenvector corresponding to the eigenvalue for which $|\lambda| = \rho(\mathbf{H})$ holds. Thus, one has $\ln\|\Delta w_n\|_2 = \ln\|\lambda^n\|_2 + \ln\|\xi\| = n\rho(\mathbf{H}) + \ln\|\Delta w_0\|_2$.
\par It should be noted that in Eq. (\ref{equ:method:6}), $\lambda_i^{\pm}$ is divided into root terms and non-root terms. Considering that $A^TA$ is Hermitian, all of their eigenvalues are real numbers. Therefore, whether the internal expression of the root term is positive or negative directly determines whether it contributes to the imaginary or real part of the eigenvalue. Based on the condition Eq. (\ref{equ:method:2}), we can derive the following inequality:
\begin{equation*}
    \begin{aligned}
        (2\hat{L}+2\hat{\mu}-|\sigma_i|^2)\le 4(\hat{L}-\hat{\mu})^2
    \end{aligned}
\end{equation*}
where the equality holds since $|\sigma_i|^2\in[\mu,L]\subseteq[\hat{\mu},\hat{L}]$. Based on this series of inequalities, we can conclude that the root term in Eq. (\ref{equ:method:6}) corresponds to the imaginary part, while the remaining terms constitute the real part. This allows us to derive the value of $\rho(\mathbf{H})$ as follows:
\begin{equation}
    \begin{aligned}
        \label{equ:method:9}
        \rho(\mathbf{H})&=\sqrt{\beta}=1-\frac{2}{\hat{\kappa}+1},
    \end{aligned}
\end{equation}
Finally, by applying the inequality $\ln(1+x) \le x$ to Eq. (\ref{equ:method:9}) and substituting it into Eq. (\ref{equ:method:8}), we obtain the following sufficient condition for convergence step to ensure the iteration termination condition:
\begin{equation}
    \begin{aligned}
        \label{equ:method:10}
        T&\gtrsim \ln\delta^{-1}\cdot \hat{\kappa}.
    \end{aligned}
\end{equation}
The result of Eq. (\ref{equ:method:10}) indicates that under our parameter settings, the convergence step of the MAG method is of the same order as both $\hat{\kappa}$ and the condition number $\kappa$ of matrix $A$. This represents a significant improvement over the quadratic order of gradient methods, as will be discussed in detail later.

\subsection{Effect of Auxiliary Variables on Convergence}

\label{section:relative}
\par When determining the convergence step time, we employed the error term $\Delta w_n=w_n-w_\infty$ proposed in Eq. (\ref{equ:method:4}). In practice, if the magnitude of a certain element in $\Delta w_n$ far exceeds that of the other elements, the error calculation becomes dominated by this single value, diminishing the contributions of the remaining elements. Consequently, the relative convergence of our algorithm may be called into question.
\par To address this, we adopt a relative error analysis to investigate whether the MAG method ensures that all elements hold equal significance in the error analysis. Let the elements of the vector $\Delta \hat{w}_n$ be defined such that $(\Delta \hat{w}_n)_i=\frac{(\Delta w_n)_i}{(w_\infty)_i}$, based on the formulation in Eq. (\ref{equ:method:4}), the relative error satisfies the following ODE:
\begin{equation*}
    \begin{aligned}
        \Delta \hat{w}_{n+1}=\hat{\mathbf{H}}\Delta \hat{w}_n,
    \end{aligned}
\end{equation*}
where $\hat{\mathbf{H}}=\diag(w_\infty)^{-1}\mathbf{H}\diag(w_\infty)$. Using the same analytical approach as in Eq. (\ref{equ:method:7}), we can derive the following inequality:
\begin{equation*}
    \begin{aligned}
        \ln\frac{\|\Delta \hat{w}_n\|_2}{\|\Delta \hat{w}_0\|_2}
        %&\le n\ln\rho(\hat{\mathbf{H}})+o(1)
        &=n\ln\rho(\mathbf{H})+\ln\kappa_2(w_\infty)+o(1),
    \end{aligned}
\end{equation*}
where $\kappa_2(w_\infty)=\frac{\max\limits_{i=1}^{2n}(w_\infty)_i}{\min\limits_{i=1}^{2n}(w_\infty)_i}$ is the condition number of $w_\infty$. Here, we utilize the fact that similarity transformations do not alter the eigenvalues of a matrix, thereby preserving the spectral radius, i.e., $\rho(\mathbf{H})=\rho(\hat{\mathbf{H}})$.
\par In the MAG method, the auxiliary variable we introduced is the look-ahead variable. The magnitude of this variable is not significantly different from that of the solution variable, so one only needs to ensure that the condition number of $u_\infty$ is not large. Therefore, under the condition that the estimated convergence step $T$ remains unchanged, the number of iterations obtained in Eq. (\ref{equ:method:10}) can still satisfy the requirement for relative convergence with small condition number $\kappa(u_\infty)$. This indicates that if there are not significant disparities in the element values of $u$, each element's contribution to convergence is almost same in our method. This constitutes the key advantage of our approach over the damped method proposed by Gu {\it et al.}, the detailed comparison will be provided in the following section.
\subsection{Comparision with Existed Methods}
\par In this section, we will compare our method with existing algorithms that use Hamiltonian simulation to solve linear systems $Au = b$, focusing primarily on two methods: the gradient descent method and the damped dynamical systems method.
\subsubsection{Gradient Descent Method}
\label{section:gradient}
\par The gradient method is essentially derived from solving the optimization problem $\|Ax - b\|_2$, leading to the following gradient flow ODE:
\begin{equation}
    \begin{aligned}
        \label{equ:compare:1}
        \frac{du(t)}{dt}=A^Tb-A^TAu,
    \end{aligned}
\end{equation}
in which the invertibility of $A$ and the positive definiteness of $A^TA$ ensure the existence of a stable solution for this ODE. Note that the ODE presented in Eq. (\ref{equ:compare:1}) is a special case of the models in Hu {\it et al.} \cite{Hu2024QuantumMultiscale}, Jin {\it et al.} \cite{Jin2025Precondition}, Hu {\it et al.} \cite{Hu2025preprint}, and Yang {\it et al.} \cite{Yang2025Linear}, as their original theories do not cover all possible cases. Therefore, we will not provide their detailed results here. Directly applying the error analysis method proposed by Hu {\it et al.} \cite{Hu2024QuantumMultiscale} and letting $\Delta u(t)=u(t)-u_\infty$ with $\Delta u_0=\Delta u(0)$, we obtain:
\begin{equation*}
    \begin{aligned}
        \|\Delta u(t)\|_2&\le e^{-\sigma_{\min}^2 t}\|\Delta u_0\|_2.
    \end{aligned}
\end{equation*}
Using the same evolution time calculation method as in Eq. (\ref{equ:method:7}), we define $T$ as the time required to achieve a global error smaller than $\delta$, yielding:
\begin{equation*}
    \begin{aligned}
        T\gtrsim \frac{\ln\delta^{-1}}{\sigma_{\min}^2}.
    \end{aligned}
\end{equation*}
When $\sigma_{\max}$ is of constant order. Using an analysis similar to that of Eq. (\ref{equ:schro:6}), we may assume that $\|A\|_{\max}$ and $\|b\|_{\max}$ are of the same order of magnitude, thereby obtaining the query complexity of the gradient descent method as
\begin{equation}
    \begin{aligned}
        \label{equ:compare:2}
        \mathcal{Q}=\tilde{\mathcal{O}}(\ln\delta^{-1}\log N_p s^2\kappa_g\ln \kappa_g),
    \end{aligned}
\end{equation}
where $\kappa_g=\frac{\|A^TA\|_{\max}}{\sigma_{\min}^2}$. This results in a squared multiple of our evolution time $\hat{\kappa}$, and this also explains the significant speed improvement we mentioned earlier compared to the gradient descent method.
\subsubsection{Damped Dynamical Systems Method}
\label{section:damped}
\par Recently, Gu {\it et al.} \cite{Gu2025Helmholtz} improved the gradient descent method by replacing the first-order derivative in Eq. (\ref{equ:compare:1}) with a damped second-order derivative, whose specific form is as follows \cite{Sandin2016Damped,Gulliksson2017TheDD,Ogren2020Damped,Gulliksson2021Damped}:
\begin{equation}
    \begin{aligned}
        \label{equ:compare:3}
        \frac{d^2u(t)}{dt^2}+\gamma\frac{du(t)}{dt}=A^Tb-A^TAu,
    \end{aligned}
\end{equation}
where $\gamma$ must satisfy the condition $\gamma < 2\sigma_{\min}$, this equation has been extensively studied in previous works \cite{Alvarez2000Damped,Begout2015Damped}. Under such conditions, we can easily derive the following global error relationship under the 2-norm
\begin{equation}
    \begin{aligned}
        \|\Delta u(t)\|_2&\le e^{-\frac{\gamma}{2} t}\|\Delta u_0\|_2.
    \end{aligned}
\end{equation}
Similarly, we can calculate the lower bound of the evolution time $T$ that satisfies the convergence condition:
\begin{equation*}
    \begin{aligned}
        T\gtrsim \frac{\ln\delta^{-1}}{\sigma_{\min}},
    \end{aligned}
\end{equation*}
We can assumed $\frac{\d u(t)}{\d t}=-A^Tv(t)$ and $w(t)=[u(t);v(t)]$, and provided an equivalent ODE:
\begin{equation}
    \begin{aligned}
        \label{equ:compare:5}
        \frac{\d w(t)}{\d t}=Jw(t)+G.
    \end{aligned}
\end{equation}
in which $J$ and $G$ are defined as
\begin{equation*}
    \begin{aligned}
        J=\begin{bmatrix}
            O & -A^T\\
            A & -\gamma I
        \end{bmatrix},\ 
        G=\begin{bmatrix}
            0\\
            -b
        \end{bmatrix}.
    \end{aligned}
\end{equation*}
Applying a method similar to that used in Eq. (\ref{equ:schro:6}), we can derive the query complexity for the damped dynamical system method:
\begin{equation}
    \begin{aligned}
        \label{equ:compare:4}
        \mathcal{Q}=\tilde{\mathcal{O}}(\ln\delta^{-1}\log N_p s^2\kappa_d\ln \kappa_d),
    \end{aligned}
\end{equation}
where $\kappa_d = \frac{\|A\|{\max}}{\sigma{\min}}$. This result matches the order of magnitude of our derived bound, indicating that the query complexities of the two methods are comparable.
\subsubsection{Advantages of Momentum Accelerated Gradient}
\par To compare the differences between our method and the damped method, we select two experiments here to illustrate from two perspectives. First, both methods utilize auxiliary variables, and when estimating the global error, both consider the norm of the concatenated vector of the solved variable $u$ and the auxiliary variable $v$. This leads to a situation where, under the condition of setting the same error tolerance $\delta$, the ratio between the auxiliary variable and the solved variable can significantly impact the accuracy of the solved variable $u$. Here, we select an example where $Au = b$, i.e., $A = \text{diag}([10,0.1])$ and $b = [1;1]$, to illustrate this issue. Although we have chosen a simple case, we can extend the comparison to all scenarios through spectral decomposition of diagonal and symmetric matrices. However, we will not delve into the details here.
\par
\begin{figure}[htbp]
    \centering
    \subfloat[]{
      \includegraphics[width=0.49\linewidth]{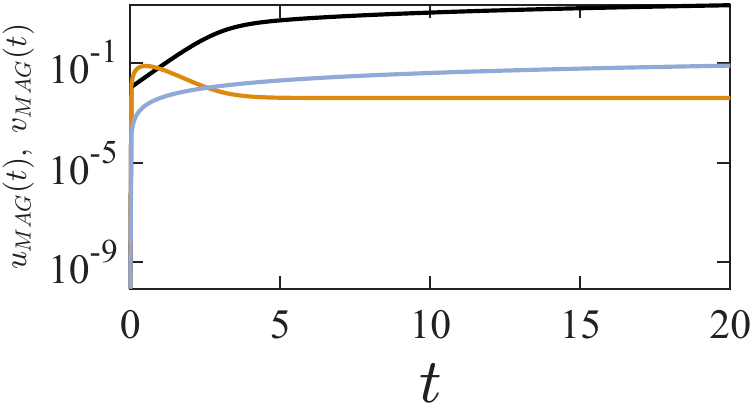}
    }
    \subfloat[]{
      \includegraphics[width=0.49\linewidth]{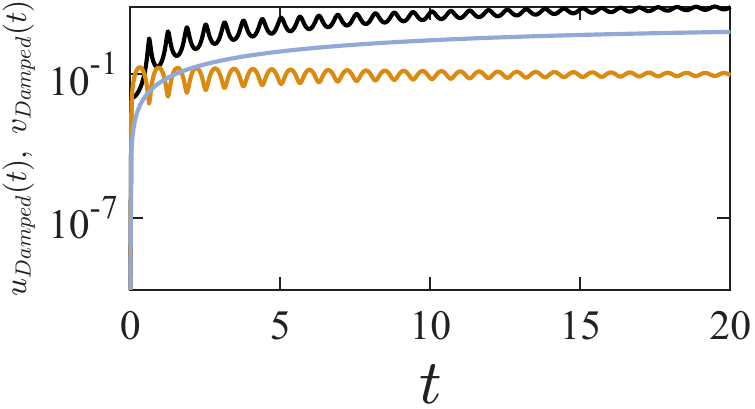}
    }
    \caption{\textbf{Comparison of the Solved Variable and the Auxiliary Variable.} The solved variable (blue line), auxiliary variable (orange line), and their ratio (auxiliary/solved, black dotted line) are shown. (a) the MAG method. (b) the dumped dynamical systems method.}
    \label{figure:1}
\end{figure}

 We solve the ODEs shown in Eq. (\ref{equ:schro:1}) and Eq. (\ref{equ:compare:5}) on the zero boundary, and the obtained results are denoted as $w_{\text{MAG}}$ and $w_{\text{Damped}}$ respectively, with their components $u$ and $v$ also using the same notation. Regarding the selection of parameters $\alpha$, $\beta$, and $\gamma$, we set the upper bound of the maximum singular value $\hat{\sigma}_{\max} = 5 \times 1.05$ and the lower bound of the minimum singular value $\hat{\sigma}_{\min} = 5 \times 0.95$. Then, we use Eq. (\ref{equ:method:2}) with $\gamma = 2\hat{\sigma}_{\min}$ to configure the parameters. As shown in Fig. \ref{figure:1}, we select the first component of $u$ for plotting. It is clearly observed that both $u$ and $v$ computed by our method converge stably, and their ratio remains relatively constant. In contrast, the damped method exhibits significant periodic discrepancies, leading to an unstable ratio that directly affects the accuracy of the solved variable $u$. This is because the auxiliary variable selected by the damped method tends to approach zero and leads to a very large $\kappa_2(w_\infty)$, making it impossible to satisfy the relative convergence condition presented in Section \ref{section:relative}.
\par
\begin{figure}[htbp]
    \centering
    \subfloat[]{
      \includegraphics[width=0.24\linewidth]{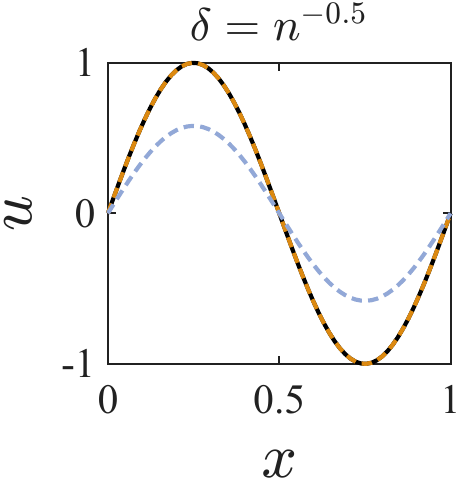}
    }
    \subfloat[]{
      \includegraphics[width=0.24\linewidth]{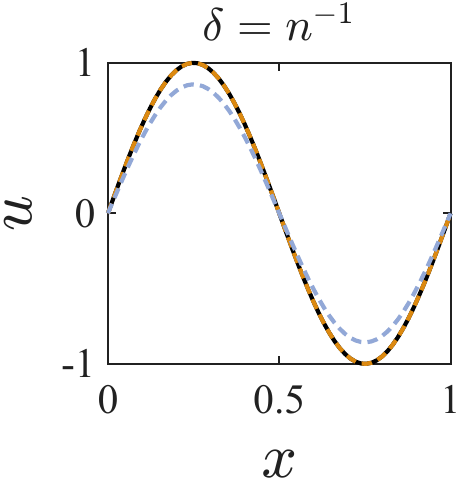}
    }
    \subfloat[]{
      \includegraphics[width=0.24\linewidth]{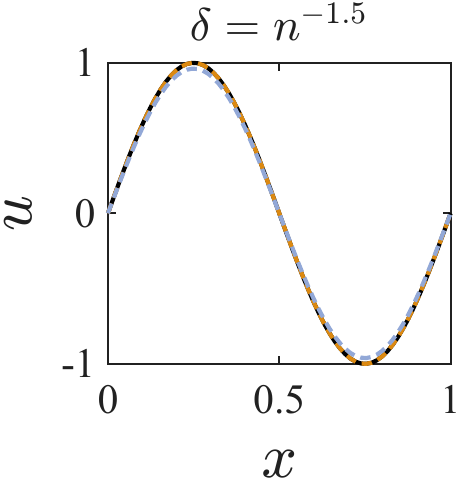}
    }
    \subfloat[]{
      \includegraphics[width=0.24\linewidth]{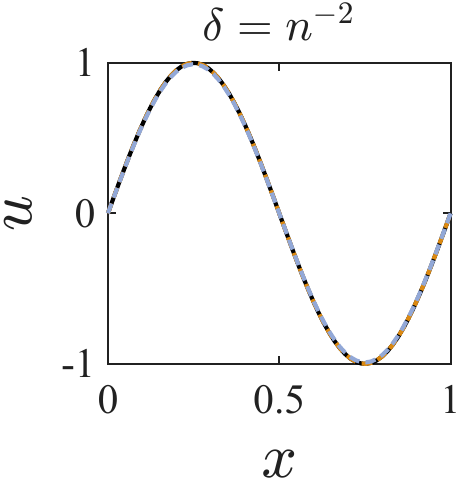}
    }
    \caption{\textbf{Solution for the Toy Problem.} Comparison of the true solution (black), results from the MAG method (orange), and results from the damped dynamical system method (blue) under different error $\delta$ selections: $\delta=n^{-0.5},n^{-1},n^{-1.5},n^{-2}$ and zero initial value..}
    \label{figure:2}
\end{figure}

 Furthermore, since our MAG method satisfies the relative convergence condition, whereas the damped method does not. This allows the value of the auxiliary variable to influence the convergence rate. To verify this, we demonstrate it through numerical experiments on the ODE problem $u(x) = f(x)$, where $x \in [0,1]$, with the boundary condition $f(x) = 2\sin(2\pi x)$. The number of discrete points is $n = 16$, and the detailed process is shown in Fig. \ref{figure:2}. It can be observed that our momentum-accelerated gradient method (orange curve) converges more readily than the damped dynamical systems method (blue curve).

\section{Quantum Framework for Solving Momentum Accelerated Gradient}

\label{section:3}

\subsection{Schr\"odingeration Method for Momentum Accelerated Gradient}

\subsubsection{Framework of Schr\"odingeration Method}

\par To simulate the Hamiltonian simulation of the iterative equation in Eq. (\ref{equ:method:4}), we first need to convert it into an ODE. Then, using the Schr\"odingerization method, we transform this ODE into an equivalent Hamiltonian operator simulation. Here, we adopt the approach proposed by Jin {\it et al.} \cite{Jin2024QuantumSO} to reformulate Eq. (\ref{equ:method:4}) into the following form:
\begin{equation}
    \begin{aligned}
        \label{equ:schro:1}
        \frac{dw(t)}{dt}=(\mathbf{H}-I)w(t)+\mathbf{F}.
    \end{aligned}
\end{equation}
Here, we employ the following transformation: Let $w_n=w(n\tau)=w(t)$, which gives $w_{n+1}-w_n=\frac{dw(t)}{dt}\tau$. By setting $\tau=1$, one obtains the ODE shown in Eq. (\ref{equ:schro:1}). Consequently, the time interval for this ODE can be constrained to $[0,T]$. Moreover, since we have already computed that $\rho(\mathbf{H})<1$, it follows that the real parts of the eigenvalues of $\mathbf{H}-I$ are all negative. This ensures the stability of solving this iterative scheme using ODE methods.
\par Next, to apply the Schr\"odingerization method \cite{Jin2023Detailes,Jin2024Schrodingerization}, we reformulate Eq. (\ref{equ:schro:1}) as a homogeneous system:
\begin{equation}
    \begin{aligned}
        \label{equ:schro:2}
        \frac{dw_{\text{homo}}(t)}{dt}=
        \mathbf{H}_{\text{homo}}
        w_{\text{homo}}(t),
    \end{aligned}
\end{equation}
where $\mathbf{H}_{\text{homo}}=\begin{bmatrix}\mathbf{H}-I & I\\\mathbf{O} & \mathbf{O}\end{bmatrix}$ and $w_{\text{homo}}(t)=[w(t);\mathbf{F}]$, with initial state $w_{\text{homo}}(0)=[w_0;\mathbf{F}]$. The matrix $\mathbf{H}_{\text{homo}}$ is separated into Hermitian $\mathbf{H}_{\text{homo,1}}=\frac{\mathbf{H}_{\text{homo}}+\mathbf{H}_{\text{homo}}^{\dagger}}{2}-I$ and anti-Hermitian $\mathbf{H}_{\text{homo,2}}=\frac{\mathbf{H}_{\text{homo}}-\mathbf{H}_{\text{homo}}^{\dagger}}{2i}$ components:
\begin{gather*}
    \mathbf{H}_{\text{homo}}=\mathbf{H}_{\text{homo,1}}+ i\mathbf{H}_{\text{homo,2}},
\end{gather*}
Using the warped phase transformation $w_{\text{warp}}(t,p)=e^{-p}w_{\text{homo}}(t)$ for $p>0$ and extending symmetrically to $p<0$, we convert Eq. (\ref{equ:schro:2}) into a convection system:
\begin{equation}
    \begin{aligned}
        \label{equ:schro:3}
        \frac{\partial w_{\text{warp}}(t,p)}{\partial t}&=(\mathbf{H}_1-i\mathbf{H_2})w_{\text{warp}}(t,p)=\mathbf{H}_1\partial_pw_{\text{warp}}(t,p)-i\mathbf{H}_2w_{\text{warp}}(t,p),\\
        w_{\text{warp}}(0,p)&=e^{-\|p\|}w_{\text{homo}}(0),
    \end{aligned}
\end{equation}
\par We then discretize $p$ via Fourier transformation over points $p_0<p_1<\cdots<p_{N_p}$, where $\Delta p=(R-L)/N_p$ and $p_k=-L+k\Delta p$. The vector $u_{\text{Four}}(t)$ is constructed as:
\begin{equation*}
    \begin{aligned}
        w_{\text{Four},i}(t)&=\sum\limits_{k\in[N_p]} w_{\text{warp},i}(t,p_k)|k\rangle,\\
        w_{\text{Four}}(t) &= [w_{\text{Four},1}(t),\cdots,w_{\text{Four},N_p}(t)]^T.
    \end{aligned}
\end{equation*}
The discretized Fourier spectrum yields:
\begin{equation}
    \begin{aligned}
        \label{equ:schro:4}
        \frac{\d w_{\text{Four}}(t)}{\d t}&=i(\mathbf{H}_1\otimes P_{\vartheta})w_{\text{Four}}(t)-i(\mathbf{H}_2\otimes I)w_{\text{Four}}(t),\\
        w_{\text{Four}}(0) &= [e^{-|p_0|},\cdots,e^{-|p_{N_p-1}|}]^T \otimes w_0.
    \end{aligned}
\end{equation}
Here, $P_{\vartheta}$ represents the momentum operator $-i\partial_p$ in matrix form. Its diagonalization $D_{\vartheta}=\phi^{-1}P_{\vartheta}\phi$ produces diagonal entries $\vartheta_{-N_p/2}$ to $\vartheta_{N_p/2-1}$, with $\phi_{j\ell}=\phi_\ell(x_j)$ where $\phi_\ell(x)=e^{i\vartheta_\ell(x-L)}$. The eigenvalues are $\vartheta_{\ell}=\pi \ell$ for $\ell=-N_p/2,...,N_p/2-1$. Through these transformations, one can obtain the ODE for $w_{\text{schr}}=(I\otimes\phi^{-1})w_{\text{Four}}$
\begin{equation}
    \begin{aligned}
        \label{equ:schro:5}
        \frac{\d w_{\text{schr}}(t)}{\d t}&=i(\mathbf{H}_1\otimes D_{\vartheta}-\mathbf{H}_2\otimes I)w_{\text{schr}}(t):=i\mathbf{H}_{\text{schr}}\cdot w_{\text{schr}}(t),\\
        w_{\text{schr}}(0)&=(\phi^{-1}\otimes I)w_{\text{Four}}(0).
    \end{aligned}
\end{equation}
\subsubsection{Restore of Solution}
\par The original solution $u(t)$ can be reconstructed through either from $w_{\text{warp}}(t,q)$:
\begin{equation*}
    \begin{aligned}
    &\text{Single-point method: }w(t)=e^{-p_k}w_{\text{warp}}(t,p_R),\\
    &\text{Integral method: }w(t)=\frac{1}{e^{p_R}-1}\int_0^{p^R} w_{\text{warp}}(t,q)\d q.
    \end{aligned}
\end{equation*}
For a more in-depth discussion on this, please refer to the improvement based on the eigenvalues of $\mathbf{H}_1$ proposed by Jin {\it et al.} \cite{Jin2025LinearSystems}.
\begin{theorem}
 \cite{Jin2025LinearSystems} Consider the case in which the dominant eigenvalue of $H_1$ (with eigenvalues ordered as $\lambda_1 \ge \lambda_2 \ge \cdots \ge \lambda_N$), denoted by $\lambda_1(H_1)$, is a positive value. Then, the solution to Eq. (\ref{equ:schro:2}) admits the representation:
\begin{equation}
    u(t)=e^pu_{\text{warp}}(t,p)\text{ for any }p>p^\Diamond=\max\{\lambda_1(H_1)t,0\},
\end{equation}
where the threshold is given by $p^\Diamond=\max\left\{\lambda_1(H_1)t,0\right\}$. An alternative formulation is:
\begin{equation}
    u(t)=e^p\int_p^\infty u_{\text{warp}}(t,q)dq\text{ for }p>p^\Diamond.
\end{equation}
\end{theorem}
\subsubsection{Query Complexity Analysis}
\par To determine the query complexity for simulating Eq. (\ref{equ:schro:5}), we utilize the classical result proposed by Berry {\it et al.} \cite{Berry2015Hamiltonian}.
\begin{lemma}
\par \cite{Berry2015Hamiltonian} For a matrix $\mathbf{H}_{\text{schr}}$ with sparsity $s(\mathbf{H}_{\text{schr}})$, simulating the Hamiltonian on $m_{\mathbf{H}_{\text{schr}}}=\mathcal{O}(\ln n)$ qubits with error $\delta$ requires queries of order
\begin{equation*}
    \begin{aligned}
        \mathcal{Q}(\mathbf{H}_{\text{schr}})=\mathcal{O}\left(\chi\frac{\ln(\chi/\delta)}{\ln\ln(\chi/\delta)}\right),
    \end{aligned}
\end{equation*}
and the number of additional 2-qubit gates of order
\begin{equation*}
    \begin{aligned}
        \mathcal{C}(\mathbf{H}_{\text{schr}})=\mathcal{O}\left(\chi[m_{\mathbf{H}_{schr}}+\ln^{2.5}(\chi/\delta)]\frac{\ln(\chi/\delta)}{\ln\ln(\chi/\delta)}\right),
    \end{aligned}
\end{equation*}
where $\chi = s(\mathbf{H}_{\text{schr}}) \|\mathbf{H}_{\text{schr}} \|_{\max} T$ and $T$ is the evolution time.
\qed
\end{lemma}
\par For the specific calculation of $\chi$, we proceed step by step. First, regarding the sparsity $s(\mathbf{H}_{\text{schr}})$, we note that $s(\mathbf{H}_{\text{schr}}) \sim s(A^TA)$, and it can be derived that $s(A^TA) = \mathcal{O}(s^2)$ with $s=s(A)$. Next, for the calculation of the max norm, since $\|\mathbf{H}_{\text{schr}} \|_{\max} \le \|\mathbf{H}_{\text{homo,1}} \|_{\max} \|D_\eta \|_{\max}$, and given that $\|\mathbf{H}_{\text{homo,1}}\|_{\max} \sim \max\{\|\alpha A^TA\|_{\max},\|\sqrt{\alpha\beta} A\|_{\max}\} = \mathcal{O}(1)$, $\|D_\eta\|_{\max}=\log N_p$, this is due to the parameter setting in Eq. (\ref{equ:method:2}), where $\alpha < \frac{1}{\sigma_{\max}^2}$. Thus, we obtain $\|\mathbf{H}_{\text{schr}}\|_{\max} = \mathcal{O}(\log N_p)$.
%({\color {green} don't understand this sentence}).
Finally, for the evolution time $T$, under the setting of Eq. (\ref{equ:schro:1}), the maximum time to solve is $T$, as given by Eq. (\ref{equ:method:10}). Combining these conditions, we conclude that $\chi = \mathcal{O}\left(\ln\delta^{-1} N_p s^2 \hat{\kappa}\right)$, and the query complexity for simulating Eq. (\ref{equ:schro:5}) under the framework of Eq. (\ref{equ:method:1}) is:  
\begin{equation}
    \begin{aligned}
        \label{equ:schro:6}
        \mathcal{Q}(\mathbf{H}_{\text{schr}})=\tilde{\mathcal{O}}\left(\ln\delta^{-1} \log N_p s^2 \hat{\kappa}\ln(\hat{\kappa})\right),
    \end{aligned}
\end{equation}
and the number of gates is
\begin{equation}
    \begin{aligned}
        \label{equ:schro:7}
        \mathcal{C}(\mathbf{H}_{\text{schr}})=\tilde{\mathcal{O}}\left(\ln\delta^{-1} \ln n\log N_p s^2 \hat{\kappa}\ln(\hat{\kappa})\right),
    \end{aligned}
\end{equation}
in which $\tilde{\mathcal{O}}$ is the order ignoring $\ln\ln$ term.
\par By comparing the result in Eq. (\ref{equ:schro:6}) with the query complexity of the HHL algorithm under error $\delta$, i.e., $\mathcal{O}(\ln\delta^{-1} s \kappa)$, we observe that, apart from the term $s$, all other factors are of the same order. This demonstrates that, under the condition of dealing with a sparse matrix $A$, our proposed method achieves the same query complexity as the HHL algorithm.

\subsection{Qubit Implementation of Momentum Accelerated Gradient}

\subsubsection{Implementation for Hamiltonian Simulation}

\label{section:hamiltonian}
\par In quantum computing, block encoding is a method for representing non-unitary matrices. Given a matrix $A$, its block encoding is a unitary matrix $U_A$ such that: $U_A = \begin{bmatrix}A / \alpha_A & * \\ * & *\end{bmatrix}$, where $ \alpha_A \ge \|A\| $ (typically $\alpha_A = \|A\|$), and $[*]$ denotes unimportant submatrices. Its detailed definition is as follows:
\begin{definition}[Block encoding]
Consider an $n$-qubit matrix $A$ and define the projection operator $\Pi = \langle 0^m | \otimes I_n$, where $I_n$ represents the $n$-qubit identity operator. We say that a $(m+n)$-qubit unitary operator $U_A$ constitutes an $(\alpha_A, m, \varepsilon)$-block-encoding of $A$ when there exist positive constants $\alpha_A$ and $\varepsilon$ satisfying the approximation condition:
\begin{equation*}
    \begin{aligned}
    \| A - \alpha_A \Pi U_A \Pi^\dagger \| = \| A - \alpha_A (\langle 0^m | \otimes I_n) U_A (| 0^m \rangle \otimes I_n) \| \leq \varepsilon.
    \end{aligned}
\end{equation*}
This formulation establishes a quantitative relationship between the target matrix $A$ and its unitary encoding $U_A$ with precision parameters $\alpha_A$ and $\varepsilon$.
\end{definition}
\noindent We need to note that the three parameters $\alpha$, $m$, and $\varepsilon$ for the same matrix $A$ satisfy different properties. Among them, $\alpha_A$ can be adjusted by changing the scale of $A$, $\varepsilon$ clearly accommodates larger values, and $m$ can accommodate larger values by adding "useless" auxiliary qubits (ancilla qubits). This is very important for the subsequent block encoding.
\par In this section, we will provide the quantum simulation for the Hamiltonian system presented in Eq. (\ref{equ:schro:4}) based on this definition:
\begin{equation*}
    \begin{aligned}
    \ket{w_{\text{Four}}(T)} = \left[\phi \otimes I\right]\cdot
    \mathcal{U}(T) \cdot \left[\phi^{-1}\otimes I\right] \ket{w_{\text{Four}}(0)},
    \end{aligned}
\end{equation*}
where $\mathcal{U}(T) = e^{-i \mathbf{H}_{\text{schr}} T}$ represents a unitary evolution operator, and the phase encoding $\phi$ (or its inverse $\phi^{-1}$) is implemented using the quantum Fourier transform (QFT) or its inverse (IQFT). To simulate the Hamiltonian evolution $\mathcal{U}(T)$, existing quantum algorithms from prior research can be employed, such as those discussed in  \cite{An2023Optimal,Jin2025Precondition,Jin2025NonUnitary}.
\par Considering that $\mathbf{H}_{\text{schr}}$ is a block-diagonal matrix composed of matrices of the form $\mathbf{H}_k := p_k \mathbf{H}_{\text{homo,1}} - \mathbf{H}_{\text{homo,2}}$, our goal is to express the evolution operator $ \mathcal{U}(T) $ as the following select oracle:
\begin{equation*}
    \begin{aligned}
	 \mathbf{H}_{\text{schr}}= \sum\limits_{k=0}^{N_p-1} \ket{k}\bra{k} \otimes \mathbf{H}_k.
    \end{aligned}
\end{equation*}
Let the unitary $V_k(T)$ be the simulation of the Hamiltonian $\mathbf{H}_k$, and we assume that the block encodings of the real part $\mathbf{H}_{\text{homo,1}}$ and the imaginary part $\mathbf{H}_{\text{homo,2}}$ are constructed separately. Let $U_{H_1}$ be an $(\alpha_{H_1}, m, \varepsilon)$-block-encoding of $\mathbf{H}_{\text{homo,1}}$, and $U_{H_2}$ be an $(\alpha_{H_2}, m, \varepsilon)$-block-encoding of $\mathbf{H}_{\text{homo,2}}$, where $\alpha_{H_1} \ge \|\mathbf{H}_{\text{homo,1}}\|$, $\alpha_{H_2} \geq \|\mathbf{H}_{\text{homo,2}}\|$, and $m$ denotes the number of ancilla qubits. Then, according to the method established by An {\it et al.} \cite{An2023Optimal}, we can construct a Hamiltonian oracle $\text{HAM}_{H_{p}}$ that satisfies
\begin{equation}
    \begin{aligned}
        (\bra{0}_{m^{'}} \otimes I) \text{HAM}_{H_p}(\ket{0}_{m^{'}} \otimes I) = \frac{\sum\limits_{k=0}^{N_p-1} \ket{k}\bra{k} \otimes \mathbf{H}_k}{\alpha_{H_1} p_{\max} + \alpha_{H_2}},
    \end{aligned}
\end{equation}
where $p_{\max}$ is the largest magnitude among all discrete Fourier coefficients ($p_{\max} = \max_k |p_k|$ for $k = 0,\cdots,N_p-1$), while $m^{'}$ (with $m^{'} > m$) represents the number of expanded ancillary qubits. Remarkably, this implementation requires just a constant number of accesses ($\mathcal{O}(1)$) to the fundamental block-encoding oracles of $\mathbf{H}_{\text{homo,1}}$ and $\mathbf{H}_{\text{homo,2}}$. Leveraging the constructed Hamiltonian oracle $\text{HAM}_{H_p}$ and using the quantum singular value transformation (QSVT), we can subsequently obtain a block-encoding of $\mathcal{U}(T)$  \cite{Gilyen2019ACM}
\begin{equation*}
    \begin{aligned}
        \text{SEL}_0 = \sum\limits_{k=0}^{N_p-1} \ket{k}\bra{k} \otimes  V_k^a(T),
    \end{aligned}
\end{equation*}
where $V_k^a(T)$ is an approximate block encoding of $V_k(T)$, satisfying $|V_k^a(T) - V_k(T)| \le \delta$. The times on oracle access to both $\mathbf{H}_{\text{homo,1}}$ and $\mathbf{H}_{\text{homo,2}}$ operators is $\mathcal{O}(\alpha_H p_{\max} T + \log\delta^{-1})$ \cite{{An2023Optimal}}, where $\alpha_H \ge \max\{\alpha_{H_1},\alpha_{H_2}\}$.
\par By implementing the block-encoding protocol on the initialized quantum state $\ket{0}{m'}\ket{w_{\text{schr}}(0)}$, we obtain the transformation:
\begin{equation*}
    \begin{aligned}
    \text{SEL}_0\ket{0}_{m^{'}}\ket{w_{\text{schr}}(0)} = \ket{0}_{m^{'}}\mathcal{U}^a(T)\ket{w_{\text{schr}}(0)} + \ket{\bot},
    \end{aligned}
\end{equation*}
where $\tilde{\mathcal{U}}(T)$ represents the approximation of the ideal unitary operator $\mathcal{U}(T)$. Notably, this quantum operation requires just one single access to the state preparation oracle $O_{w_{\text{schr}}}$ that encodes the initial condition $w_{\text{schr}}(0)$.
\par Based on the previous discussions, it can be established that one can construct a quantum operation $V_0$ satisfying the following transformation:
\begin{equation*}
    \begin{aligned}
    \ket{0^{n_a}} \ket{0^w} \quad \xrightarrow{ V_0 } \quad  \frac{1}{\eta_0} \ket{0^{n_a}} \otimes w_{\text{schr}}^{a} + \ket{\bot},
    \end{aligned}
\end{equation*}
where $w_{\text{schr}}^{a}$ represents the numerically approximated solution to $w_{\text{schr}}$, expressed as
\begin{equation}
    \begin{aligned}
    \label{equ:hamiltonian:1}
     w_{\text{schr}}^{a}(T) = \mathcal{U}^a(T)w_{\text{schr}}(0),\ \eta_0 \lesssim \triangle p\sqrt{\|w_0\|^2+T^2\|\mathbf{F}\|_1^2},
    \end{aligned}
\end{equation}
and the state $\ket{\psi^\perp}$ encompasses all components orthogonal to the desired solution subspace.
\subsubsection{Block Encoding for Sparse Matrices}
\par Existing quantum algorithms demonstrate that efficient block encodings can be derived for sparse matrices using their corresponding sparse access oracles, as established in prior works \cite{Gilyen2019ACM,Chakraborty2019ICALP,LinLin2022Notes}.
\begin{definition}[Block encoding for sparse matrices]
\par Consider an $n$-qubit sparse matrix $A = (a_{ij})$ where each row and column contains no more than $s$ non-zero elements. Suppose the maximum absolute value of its entries satisfies $\max\limits_{i,j=1}^n|a_{ij}| \le 1$, and the matrix is accessible via three quantum oracles:
\begin{gather*}
O_r\ket{l}\ket{i} = \ket{r(i,l)}\ket{i}, \quad O_c\ket{l}\ket{j} = \ket{c(j,l)}\ket{j}, \\
O_A\ket{0}\ket{i,j} = \left(a_{ij}\ket{0} + \sqrt{1-|a_{ij}|^2}\ket{1}\right)\ket{i,j},
\end{gather*}
where, $r(i,l)$ and $c(j,l)$ respectively identify the position of the $l$-th non-zero element in row $i$ and column $j$. Under these conditions, we can construct an $(s, n+1)$-block-encoding of $A$ by making one query to each oracle $O_r$, $O_c$, and $O_A$, while requiring only $\mathcal{O}(n)$ basic quantum gates and a constant number of ancillary qubits.
\end{definition}
\noindent To construct the block encodings for $\mathbf{H}_{\text{homo,1}}$ and $\mathbf{H}_{\text{homo,2}}$, we require the following definitions and lemmas from existing quantum computation literature \cite{Gilyen2019ACM,Chakraborty2019ICALP}
\begin{definition}[State preparation pair]
\par Consider a vector $y \in \mathbb{C}^n$ with $\|y\|_1 \le \beta$. The unitary operators $(P_L, P_R)$ are referred to as a $(\beta, b, \varepsilon)$-state preparation pair if they satisfy the following conditions: 
\begin{equation*}  
    \begin{aligned}  
    P_L|0\rangle^{\otimes b} = \sum\limits_{j=0}^{2^b-1} c_j |j\rangle, \quad  
    P_R|0\rangle^{\otimes b} = \sum\limits_{j=0}^{2^b-1} d_j |j\rangle,  
    \end{aligned}  
\end{equation*}
such that the weighted sum of deviations satisfies $\sum\limits_{j=0}^{m-1} \left| \beta (c_j^* d_j) - y_j \right| \leq \varepsilon$, and for all indices $j \in \{m, \ldots, 2^b-1\}$, the product $c_j^* d_j$ vanishes (i.e., $c_j^* d_j = 0$).  
\end{definition}
\begin{lemma}\label{lemma:blockencoding:1}
\par \cite{Gilyen2019ACM,Chakraborty2019ICALP} Consider two $n$-qubit matrices $A_i$ ($i=0,\cdots s-1$), each with an $(\alpha_i, m_i, \varepsilon_i)$-block encoding $U_i$ and gate complexity $T_i$. The following block encodings can be constructed:
\begin{itemize}
\item Assume each $A_k$ has an $(\alpha, m, \varepsilon_1)$-block-encoding $U_k$, and let $(P_L, P_R)$ be a $(\beta, \ell, \varepsilon_2)$-state-preparation-pair for the coefficient vector $y$. Then, the linear combination $\sum\limits_{k=0}^{s-1} y_k A_k$ has an $(\alpha\beta + \alpha, m+\ell, \alpha\varepsilon_1 + \alpha\beta \varepsilon_2)$-block-encoding, with a gate complexity of $\mathcal{O}\left(\sum\limits_{k=0}^{s-1} T_k\right)$:
\begin{gather*}
    (P_L^\dagger\otimes I_m\otimes I_s)W(P_R\otimes I_m\otimes I_s),\\
    \text{with } W=\sum\limits_{i=0}^{m-1}\ket{i}\bra{i}\otimes U_i+((I-\sum\limits_{i=0}^{m-1}\ket{i}\bra{i})\otimes I_m\otimes I_s).
\end{gather*}
\item Product of matrices ($A_1 A_2$): Has an $(\alpha_1 \alpha_2, m_1 + m_2, \alpha_1 \varepsilon_2 + \alpha_2 \varepsilon_1)$-block encoding with gate complexity $\mathcal{O}(T_1 + T_2)$: $(I_{m_2}\otimes U_1)(I_{m_1}\otimes U_2)$.
\item Tensor product ($A_1 \otimes A_2$): Has an $(\alpha_1 \alpha_2, m_1 + m_2, \alpha_1^2 \varepsilon_2 + \alpha_2^2 \varepsilon_1 + \varepsilon_1 \varepsilon_2)$-block encoding with gate complexity $\mathcal{O}(T_1 + T_2)$: $U_1\otimes U_2$.
\item Scalar multiplication ($c A_2$): If $A_1 = c$ is a scalar, the block encoding of $A_2$ induces a $(c \alpha_2, m_2, c \varepsilon_2)$-block encoding for $c A_2$.
\item Conjugate Transpose ($A_i^\dagger$): Has an $(\alpha_{A_i^\dagger}, m, \varepsilon)$-block encoding with gate $\mathcal{O}(T_i)$: $U_{A_i^\dagger}$.
\item Unitary matrix ($I$): Has a $(1, 0, 0)$-block encoding $I$. 
\end{itemize}
\end{lemma}
\par In the following, we will present the block encoding of $\mathbf{H}_{\text{homo,1}}$ and $\mathbf{H}_{\text{homo,2}}$ based on the known block encoding of $A$. Due to space limitations, we will only consider the partial implementation of $\mathbf{H}_{\text{homo,1}}$ here. By using auxiliary qubits $\ket{i}\bra{j}$ to label the rows and columns of the matrix, the decomposition of $\mathbf{H}_{\text{homo,1}}$ can be obtained as follows:
\begin{equation}
    \begin{aligned}
        \mathbf{H}_{\text{homo,1}} = \sum\limits_{i,j=0}^3 \ket{i}\bra{j} \otimes J_{ij},\\ 
    \end{aligned}
\end{equation}
where the coefficients $J_{ij}$ are defined as follows (excluding zero terms)
\begin{equation*}
    \begin{aligned}
        J_{00} &= -\alpha A^TA,\ J_{01} = -\sqrt{\alpha\beta}A^T, \\
        J_{10} &= \sqrt{\alpha\beta}A,\ J_{11} = -I\\
        J_{02} &= J_{13} = J_{20} = J_{31} = \frac{1}{2}I.
    \end{aligned}
\end{equation*}
We now employ Lemma \ref{lemma:blockencoding:1} to construct the block encoding for $\ket{0}\bra{1}\otimes J_{01}$. This particular block demonstrates universal applicability, and therefore we will not elaborate on the remaining blocks. First, we assume that matrix $A$ possesses a $(\alpha_A,m,\varepsilon)$ block-encoding denoted by $U_A$ with gate complexity $T$. Meanwhile, a $(1,1,0)$-block encoding for $\ket{0}\bra{1}$ is given by the matrix $U_c=\begin{bmatrix}0 & 1 & 0 & 0\\ 0 & 0 & 0 & 1\\ 1 & 0 & 0 & 0\\ 0 & 0 & 1 & 0\end{bmatrix}$. By combining these components, we obtain
\begin{itemize}
\item $A^\dagger A$: Has an $(\alpha_{A^\dagger}\alpha_A,2m,(\alpha_{A^\dagger}+\alpha_A)\varepsilon)$-block encoding with gate complexity $\mathcal{O}(T)$: $U_{A^\dagger}U_A$.
\item $A^\dagger$: Has an $(\alpha_{A^\dagger},m,(\alpha_{A^\dagger})\varepsilon)$-block encoding with gate complexity $\mathcal{O}(T)$: $U_{A^\dagger}$.
\end{itemize}
\subsubsection{Number of Repetitions for Measurements}
\par Based on the theoretical framework developed in Section \ref{section:hamiltonian}, the approximate solution $w_{\text{approx}}(T)$ can be obtained through quantum state measurement. When performing measurements on the quantum state described in Eq. (\ref{equ:hamiltonian:1}), the probability of observing all zeros in the first $n_a$ qubits corresponds to $\left(\frac{\|w_{\text{approx}}\|}{\eta}\right)^2$. This indicates that multiple experimental repetitions are required to enhance precision, a process that has already been addressed in numerous prior studies \cite{Jin2025Precondition,Jin2025NonUnitary,Gu2025Helmholtz}. Therefore, we will not reiterate the same procedure here. Finally, through quantum amplitude amplification techniques, the required number of measurement repetitions can be approximately reduced to
\begin{equation*}
    \begin{aligned}
        g=\mathcal{O}\left(\frac{\|w(0)\|+T\|\mathbf{F}\|_1}{\|w(T)\|}\right),
    \end{aligned}
\end{equation*}
where we can disregard $w(0)$ and treat $w(T)$ as its approximate value, $w(T) \to \mathbf{H}^{-1}\mathbf{F}$. Consequently, we have $\|w(T)\| \lesssim \|\mathbf{A}^{-1}\mathbf{b}\|$ and $\|\mathbf{F}\| \lesssim \|\mathbf{A}^T\mathbf{b}\|$. This allows us to derive an upper bound estimate for the number of repetition $g$ through results presented in Eq. (\ref{equ:method:10}):
\begin{equation}
    \begin{aligned}
        g=\mathcal{O}\left(\ln\delta^{-1}\|A\|^2\hat{\kappa}\right).
    \end{aligned}
\end{equation}
\section{Numerical Examples}
\label{section:4}
\par In this section, we will validate our results on two classic elliptic equations: the Helmholtz Equation and the Biharmonic Equation. Both of these equations can be transformed into linear systems for numerical solution.
\subsection{Helmholtz Equation}
\par The Helmholtz equation \cite{Engquist2011Helmholtz,Gander2019Helmholtz} is a PDE that describes wave phenomena and vibration problems, widely used in fields such as acoustics, electromagnetics, and quantum mechanics to analyze steady-state wave behavior. Its basic form is
\begin{equation}
    \begin{aligned}
        \label{equ:example:1}
        \nabla^2 u(x,y)+k^2u(x,y)=f(x,y),\ x,y\in[0,1],
    \end{aligned}
\end{equation}
where $u(x,y)$ represents a physical quantity (such as sound pressure or an electromagnetic field), $\nabla^2$ is the Laplace operator, and $k$ is the wavenumber. Specially, the general form of the one-dimensional case is given by
\begin{equation}
    \begin{aligned}
        \label{equ:example:2}
        \partial_{xx}u(x) + k^2u(x) = f(x),\ x\in[0,1].
    \end{aligned}
\end{equation}

\subsubsection{Finite Difference Schemes}

\par First, we consider the numerical discretization for the one-dimensional Helmholtz equation subject to zero boundary condition. The domain $[0,1]$ along the $x-$axis is partitioned using a uniform grid with a spacing of $h$, generated by introducing
{$n-2$ interior nodes}. The solution and the source/forcing function are represented by the discrete vectors $\mathbf{u} = [u_1; u_2; \cdots; u_{n-1}]$ and $\mathbf{f}=[f_1;f_2;\cdots;f_{n-1}]$, respectively. Applying a finite difference approximation to the Laplacian in Equation (\ref{equ:example:2}) leads to the resulting system of linear equations:
\begin{equation*}
    \begin{aligned}
        \mathbf{A}u=b,
    \end{aligned}
\end{equation*}
in which $\mathbf{H}$ and $\mathbf{b}$ are defined as
\begin{equation*}
    \begin{aligned}
        \mathbf{A}=L_h+k^2h^2I_{n},\ \mathbf{b}=h^2\mathbf{f},
    \end{aligned}
\end{equation*}
where $L_h$ is the second-order derivative difference matrix and
\begin{equation*}
    \begin{aligned}
        L_h=
        \begin{bmatrix}
        -2 & 1 &&&\\
        1 & -2 & 1 &&\\
        & \ddots & \ddots & \ddots &\\
        && 1 & -2 & 1 \\
        &&& 1 & -2
        \end{bmatrix}.
    \end{aligned}
\end{equation*}
Furthermore, for the Robin boundary condition $\frac{\d u(0)}{\d t}=u(0)$, we can extend $\mathbf{u}$ by one term to obtain $\tilde{\mathbf{u}} = [u_0, u_1, u_2, \cdots, u_{n-1}]$, and extend $\mathbf{A}$ and $\mathbf{b}$ to the following format:
\begin{equation*}
    \begin{aligned}
        \tilde{\mathbf{A}}=\begin{bmatrix}
            -(1+2h) & \mathbf{I}_1^T\\
            \mathbf{I}_1 & \mathbf{A}
        \end{bmatrix},\ 
        \tilde{\mathbf{b}}=\begin{bmatrix}
            0\\
            \mathbf{b}
        \end{bmatrix},
    \end{aligned}
\end{equation*}
in which $\mathbf{I}_1$ is the first column of the identity matrix of size $n-2$.
%({\color {green} what is $Nx$?}).

\par {Next, we describe the discretization approach for the Helmholtz equation under zero boundary conditions. The domain $[0,1] \times [0,1]$ is discretized uniformly in both the $x$ and $y$ directions. We introduce $n-2$ interior points in each dimension, and denote the grid spacing by $h$. Let the discrete values of the solution and the source term be represented by matrices $\vec{u}$ and $\vec{f}$, respectively, where $(\vec{u})_{ij} = u(x_i, y_j)$ and $(\vec{f})_{ij} = f(x_i, y_j),(i, j = 1, 2, \cdots, n).$ We then form the corresponding vector representations through the vectorization operation $\text{vec}$, which stacks all columns of the matrix into a single column vector: $\mathbf{u} = \text{vec}(\vec{u})$ and $\mathbf{f} = \text{vec}(\vec{f})$. Using a finite difference discretization of the Laplace operator in Eq. (\ref{equ:example:1}), we obtain the linear system:}
%Next, we consider the discretization scheme for the Helmholtz equation with zero boundary conditions. We discretize the intervals $[0,1]$ in both the $x$ and $y$ directions by inserting $n-2$ points and let $h$ be the step size, and define the discrete matrices $\vec{u}$ and $\vec{f}$ with entries $(\vec{u})_{ij} = u_{ij}$ and $(\vec{f})_{ij} = f_{ij}$, where $i,j = 1, \cdots, n+1$. Further, we define $u = \text{vec}[\vec{u}]$ and $f=\text{vec}[\vec{f}]$, where $\text{vec}$ denotes the matrix vectorization operation ({\color {green} not clear what this is}). Then, we can discretize the Laplace operator in Eq. (\ref{equ:example:1}) and obtain the linear system with
\begin{equation*}
    \begin{aligned}
        \mathbf{A}=(I_n\otimes L_h+L_h\otimes I_n)+k^2h^2I_{n^2},\ \mathbf{b}=h^2f.
    \end{aligned}
\end{equation*}
{For the discretization scheme of Robin conditions in the two-dimensional equation, we can perform a tensor product extension based on the one-dimensional format and determine the boundaries,}
%({\color {green} you mean tensor product extension of the one-dimensional case?}),
and it will not be detailed provided here.

\subsubsection{Simulation}
\par
\begin{figure}[htbp]
    \centering
    \subfloat[]{\vspace{0.2em}
      \includegraphics[width=0.25\linewidth]{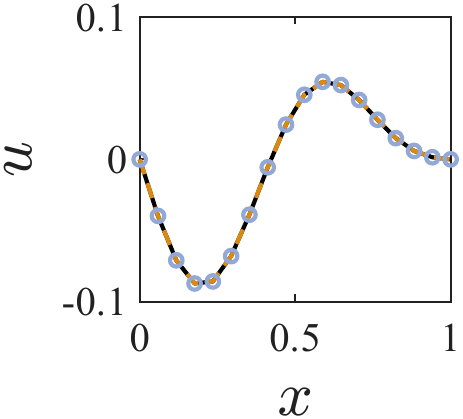}
    }
    \subfloat[]{
      \includegraphics[width=0.25\linewidth]{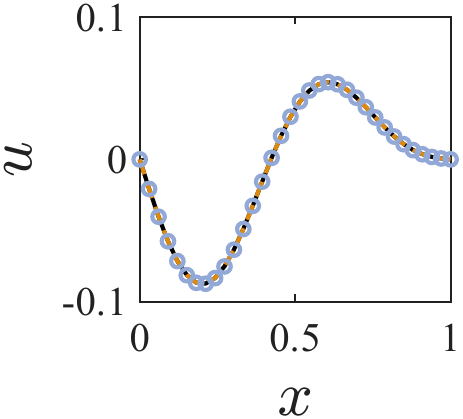}
    }
    \subfloat[]{
      \includegraphics[width=0.25\linewidth]{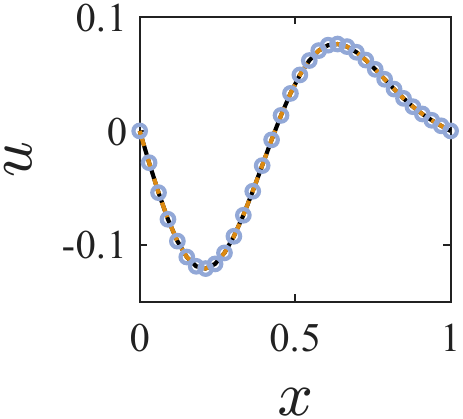}
    }\\
    \subfloat[]{\vspace{0.2em}
      \includegraphics[width=0.25\linewidth]{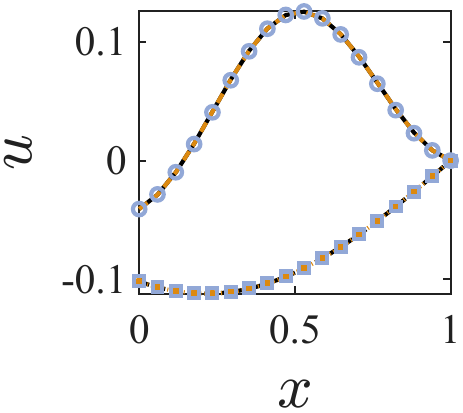}
    }
    \subfloat[]{
      \includegraphics[width=0.25\linewidth]{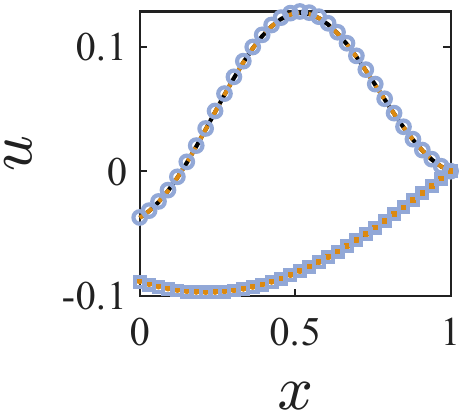}
    }
    \subfloat[]{
      \includegraphics[width=0.25\linewidth]{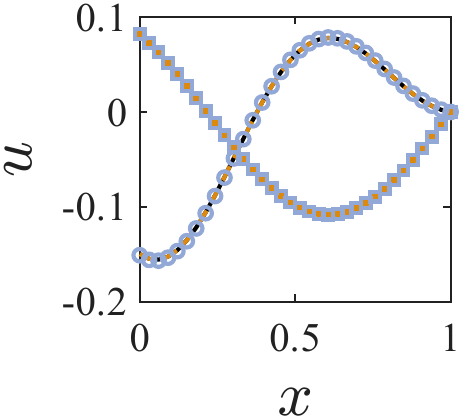}
    }
    \caption{\textbf{Solution for the One-Dimensional Helmholtz Equation.} Comparison of the true solution (black), results from the MAG method (orange), and results after Schr\"odingeration (blue). The solid line and the circle represent the real part, while the dashed line and the square represent the imaginary part. (a)-(c) with $f(x)=2\sin(2\pi x)+3\sin(3\pi x)$ and zero boundary: (a) $n=16$ and $k=2$. (b) $n=32$ and $k=2$. (c) $n=32$ and $k=4$. (d)-(f) with $f(x)=2\cos(2\pi x)$ and Robin boundary $\frac{\d u(0)}{\d x}=2iu(0)$: (d) $n=16$ and $k=2$. (e) $n=32$ and $k=2$. (f) $n=32$ and $k=4$.}
    \label{figure:3}
\end{figure}

\begin{figure}[htbp]
    \centering
    \subfloat[]{\vspace{0.2em}
      \includegraphics[width=0.25\linewidth]{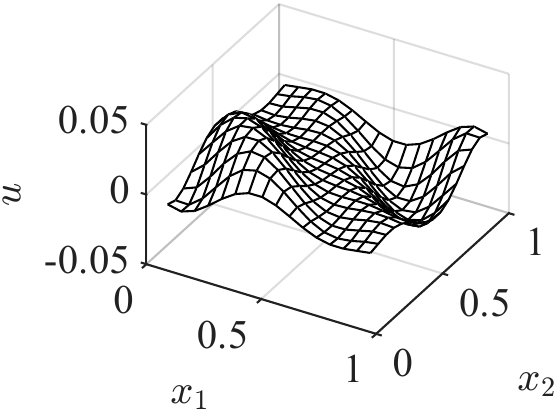}
    }
    \subfloat[]{
      \includegraphics[width=0.25\linewidth]{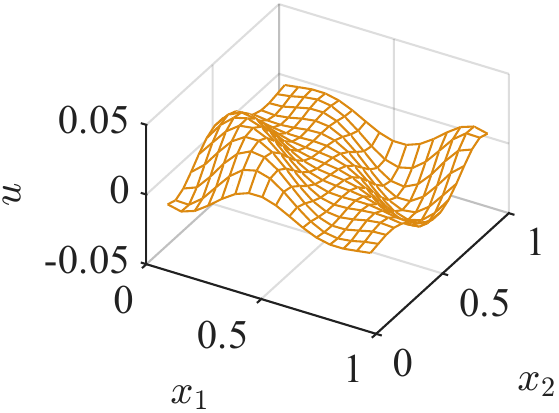}
    }
    \subfloat[]{
      \includegraphics[width=0.25\linewidth]{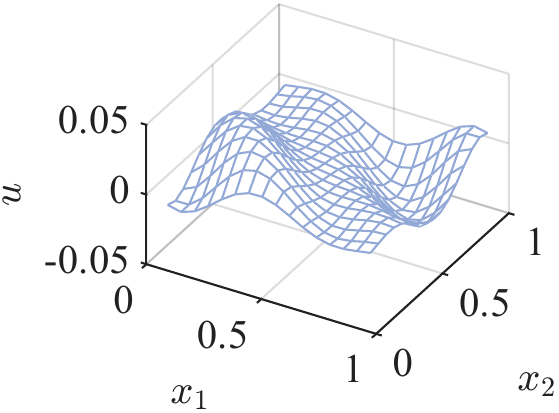}
    }\\
    \subfloat[]{\vspace{0.2em}
      \begin{minipage}{0.25\textwidth}
      \includegraphics[width=\linewidth]{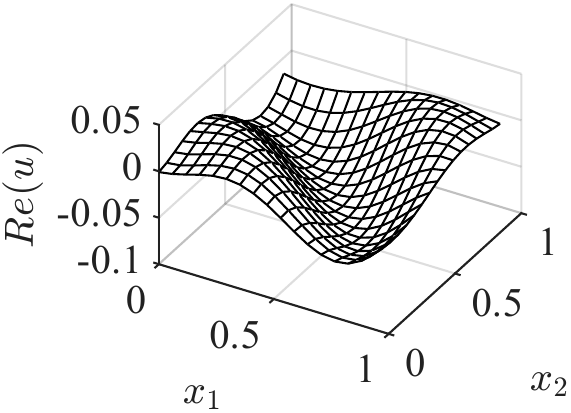}\\
      \includegraphics[width=\linewidth]{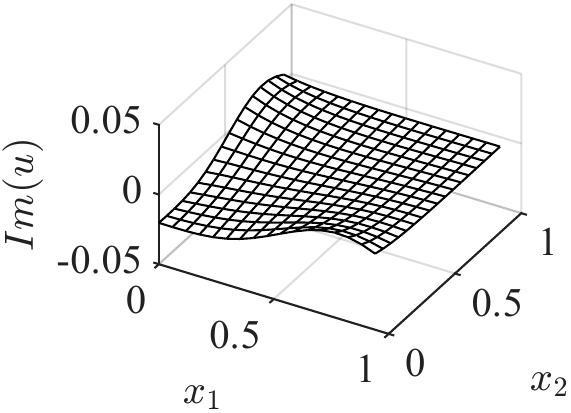}
      \end{minipage}
    }
    \subfloat[]{
      \begin{minipage}{0.25\textwidth}
      \includegraphics[width=\linewidth]{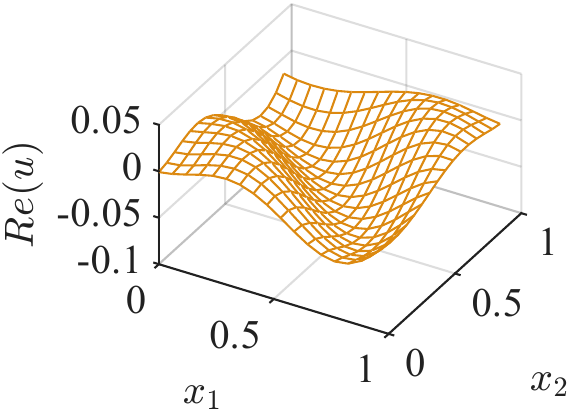}\\
      \includegraphics[width=\linewidth]{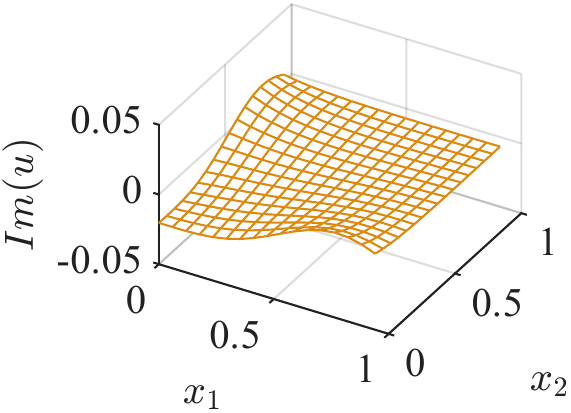}
      \end{minipage}
    }
    \subfloat[]{
      \begin{minipage}{0.25\textwidth}
      \includegraphics[width=\linewidth]{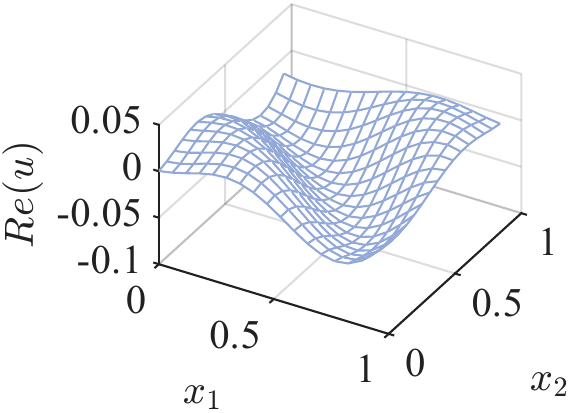}\\
      \includegraphics[width=\linewidth]{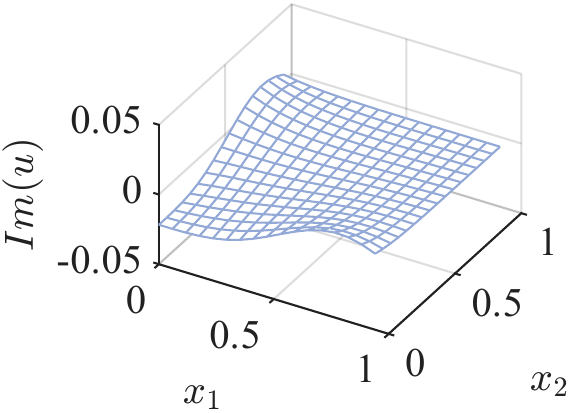}
      \end{minipage}
    }
    \caption{\textbf{Solution for the Two-Dimensional Helmholtz Equation.} with $n_x=n_y=16$ and $k=1$. (a)-(c) $f(x)=2\sin(2\pi (x+y))+3\sin(3\pi (x+y))$ and zero boundary. (d)-(f) $f(x)=2\cos(2\pi (x+y))$ and Robin boundary $\frac{\partial u(0,y)}{\partial x}=2iu(0,y)$, $\frac{\partial u(x,0)}{\partial y}=2iu(x,0)$. The first row is the real part while the second row is the imagine part.}
    \label{figure:4}
\end{figure}

 To validate our theory, we conducted numerical simulations on both one and two-dimensional Helmholtz equations, with zero boundary and Robin boundary conditions, respectively. The simulation results are presented in Figs. \ref{figure:3} and \ref{figure:4}, where it can be observed that the theoretical predictions, the results obtained via the MAG method, and the results after Schr\"odingerization all align well with each other.
\subsection{Biharmonic Equation}
\par The biharmonic equation \cite{Falk1978Biharmonic,Behrens2011Biharmonic} is a fourth-order partial differential equation that describes problems such as thin plate bending, elasticity, and fluid flow. It is widely used in structural mechanics and materials science to analyze deformation and stability of objects. Its basic form is
\begin{equation}
    \begin{aligned}
        \label{equ:example:4}
        \nabla^4 u(x,y)=f(x,y),\ x,y\in[0,1],
    \end{aligned}
\end{equation}
where $\nabla^4$ is the biharmonic operator. More generally, we can solve the following equivalent system of differential equations:
\begin{equation}
    \begin{aligned}
        \label{equ:example:3}
        \nabla^2 u(x,y)&=v(x,y),\\
        \nabla^2 v(x,y)&=f(x,y),\ x,y\in[0,1].
    \end{aligned}
\end{equation}
\subsubsection{Finite Difference Schemes}
\label{section:bi}
\par We can adopt similar notation and matrix definitions as used in the discretization of the one-dimensional Helmholtz Equation. Let $u=[u_{n-1};\cdots;u_1]$ and $v=[v_{n-1};\cdots;v_1]$. Further, let $w=[u;v]$. Then, if we impose the zero boundary conditions $u(0)=u(1)=\frac{\d^2 u(0)}{\d x^2}=\frac{\d^2 u(1)}{\d x^2}=0$, which is equivalent to $u(0)=v(0)=u(1)=v(1)=0$ (where $v(x) = \frac{\d^2 u(x)}{\d x^2}$), we obtain the following linear system:
\begin{equation*}
    \begin{aligned}
        \mathbf{A}u=b,
    \end{aligned}
\end{equation*}
in which $\mathbf{A}$ is defined as
\begin{equation*}
    \begin{aligned}
        \mathbf{A}=\begin{bmatrix}
            L_h & -h^2I_n\\
            O_{n} & L_h
        \end{bmatrix}
    \end{aligned},\ 
    \mathbf{b}=\begin{bmatrix}
        0\\
        h^2\mathbf{f}
    \end{bmatrix}.
\end{equation*}
For another boundary condition: $u(0)=u(1)=\frac{\d u(1)}{\d t}=0$, $\frac{\d u(0)}{\d t}=2$, we only need to make some adjustments to $\mathbf{b}$ by subtracting $2$ from the $n+1$-st value.
\par {Building upon the previously established notation and matrix definitions from the one-dimensional case, we now consider the two-dimensional Helmholtz equation with zero boundary conditions. Let the matrices $\vec{u}$ and $\vec{v}$ represent discrete functions on the grid, with entries defined as  $(\vec{u})_{ij} = u(x_i,y_j)$ and $(\vec{v})_{ij} = v(x_i,y_j),( i, j = 1, 2, \cdots, n+1).$ We then convert these matrices into column vectors using the vectorization operation $\text{vec}$:  $\mathbf{u} = \text{vec}(\vec{u})$ and $\mathbf{v} = \text{vec}(\vec{v})$, and define the combined vector $\mathbf{w} = [\mathbf{u}; \mathbf{v}]$. Using these representations, we derive the following definitions for the matrix $\mathbf{A}$ and the vector $\mathbf{b}$:}
%For the two-dimensional Helmholtz equation with zero boundary conditions, we adopt the notation and matrix definitions established for the one-dimensional case. Let $(\vec{u})_{ij} = u_{ij}$ and $(\vec{v})_{ij} = v_{ij}$ for $i, j = 1, \cdots, n+1$. We then define $u = \text{vec}[\vec{u}]$, $v = \text{vec}[\vec{v}]$, and $w = [u; v]$. This allows us to derive the following definitions of $\mathbf{A}$ and $\mathbf{b}$:
\begin{equation*}
    \begin{aligned}
        \mathbf{A}=\begin{bmatrix}
            I_n\otimes L_h+L_h\otimes I_n & -h^2I_{n^2}\\
            O_{n^2} & I_n\otimes L_h+L_h\otimes I_n
        \end{bmatrix}
    \end{aligned},\ 
    \mathbf{b}=\begin{bmatrix}
        0\\
        h^2\mathbf{f}
    \end{bmatrix}.
\end{equation*}
For the discrete scheme of non-zero boundaries in the two-dimensional case, it can be derived by modifying this scheme, which we will not elaborate on here.
\subsubsection{Simulation}
\par
\begin{figure}[htbp]
    \centering
    \subfloat[]{
      \includegraphics[width=0.24\linewidth]{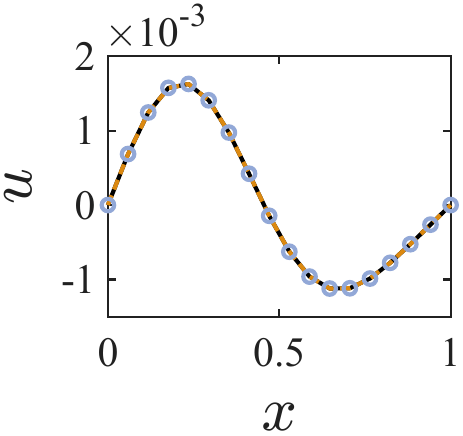}
    }
    \subfloat[]{
      \includegraphics[width=0.24\linewidth]{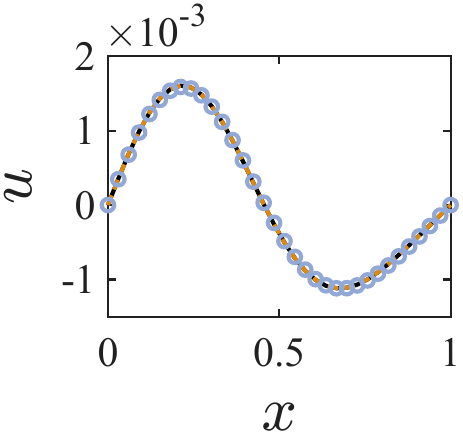}
    }
    \subfloat[]{
      \includegraphics[width=0.24\linewidth]{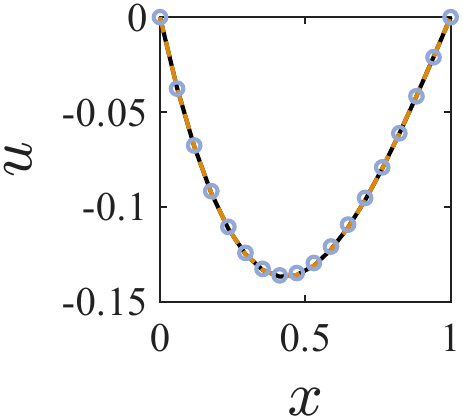}
    }
    \subfloat[]{
      \includegraphics[width=0.24\linewidth]{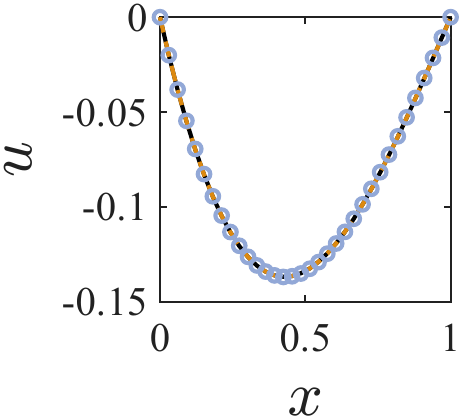}
    }
    \caption{\textbf{Solution for the One-Dimensional Biharmonic Equation.} (a)-(b) with $f(x)=2\sin(2\pi x)+3\sin(3\pi x)$ and zero boundary: (a) $n=16$. (b) $n=32$. (c)-(d) with $f(x)=2\cos(2\pi x)$ and boundary $\frac{\d^2 u(0)}{\d x^2}=2$, $u(0)=u(1)=\frac{\d^2 u(1)}{\d x^2}=0$: (c) $n=16$. (d) $n=32$.}
    \label{figure:5}
\end{figure}

\begin{figure}[htbp]
    \centering
    \subfloat[]{
      \includegraphics[width=0.25\linewidth]{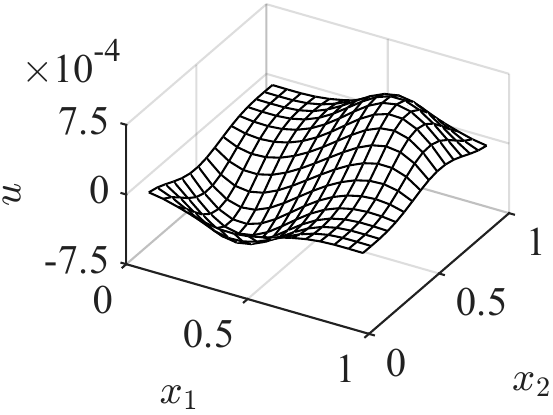}
    }
    \subfloat[]{
      \includegraphics[width=0.25\linewidth]{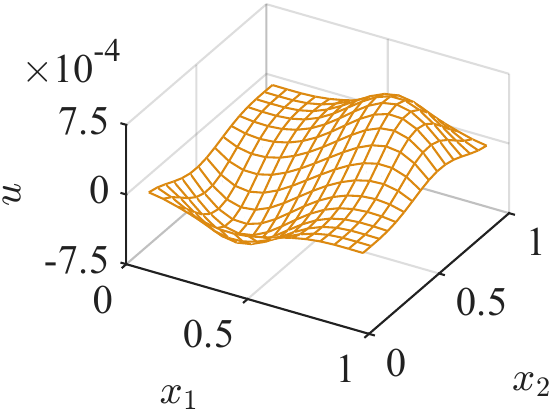}
    }
    \subfloat[]{
      \includegraphics[width=0.25\linewidth]{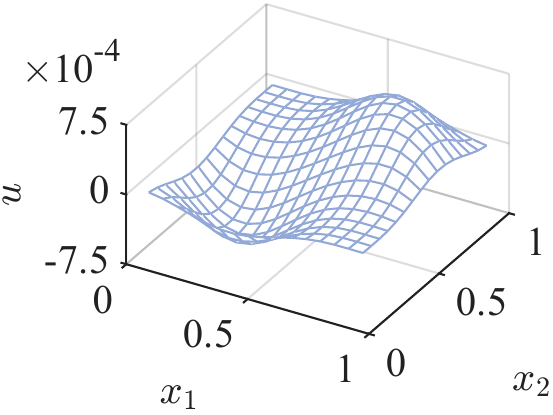}
    }\\
    \subfloat[]{
      \includegraphics[width=0.25\linewidth]{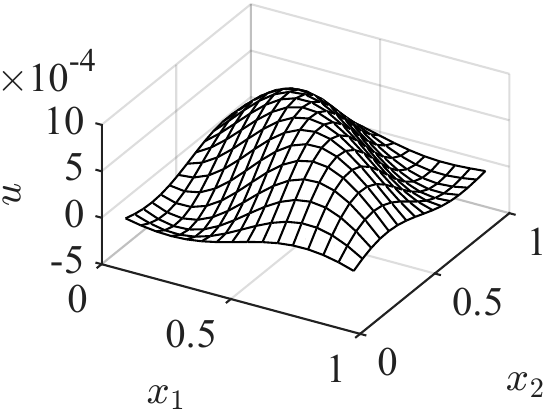}
    }
    \subfloat[]{
      \includegraphics[width=0.25\linewidth]{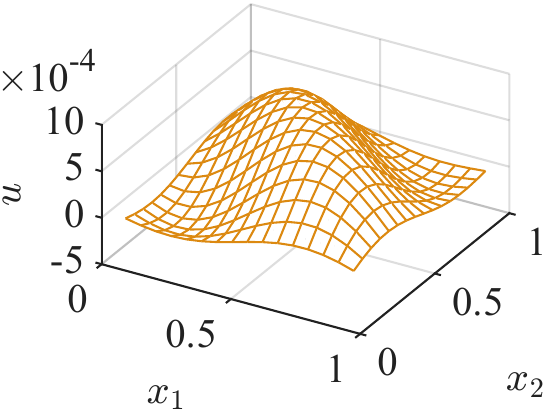}
    }
    \subfloat[]{
      \includegraphics[width=0.25\linewidth]{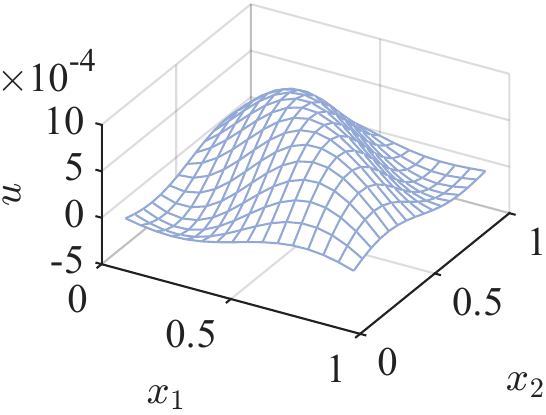}
    }
    \caption{\textbf{Solution for the Two-Dimensional Biharmonic Equation.} with $n_x=n_y=16$ and $k=1$. (a)-(c) $f(x)=2\sin(2\pi (x+y))+3\sin(3\pi (x+y))$ and zero boundary. (d)-(f) $f(x)=2\cos(2\pi (x+y))$ and boundary $\frac{\partial^2 u(0,y)}{\partial x^2}=\frac{\partial^2 u(0,y)}{\partial y^2}=2$, $u(0,y)=u(x,0)=u(1,y)=u(x,1)=\frac{\partial^2 u(1,y)}{\partial x^2}=\frac{\partial^2 u(1,y)}{\partial y^2}=0$.}
    \label{figure:6}
\end{figure}

 Similarly, we conduct numerical experiments
{under the boundary conditions as we discussed in Section \ref{section:bi}
%({\color {green} Here we have fourth order PDE so need four BCs})
%boundary conditions as the Helmholtz Equation
for both one-dimensional and two-dimensional cases}, with the results presented in Figs. \ref{figure:5} and \ref{figure:6}. It can be observed that the three results align well, demonstrating the feasibility of our theory.

\section{Conclusion and Discussion}

\par In this paper, we propose a quantum algorithm based on the momentum accelerated gradient method and the Schr\"odingerization approach. This method accelerates iteration by incorporating look-ahead points. Through theoretical analysis, we demonstrate that our algorithm achieves polynomial speedup compared to classical gradient descent methods, with query complexity comparable to the Schr\"odingerization based damped dynamical system method.
{Moreover, compared to quantum algorithms not based on Schr\"odingerization method, such as the HHL algorithm and its improvements, our method does not impose excessive demands on quantum hardware, thus enabling easier practical implementation.}
%({\color {green} still don't see advantage over HHL. Only see advantage over damped method.})
Furthermore, our analysis reveals that our method exhibits superior convergence properties to the damped method. Notably, we ensure convergence for both the solved variables and the auxiliary variables, which is significantly important for rapidly solving linear systems. On the contrary, the damped system exhibits varying proportions of the two variables at different times, and there are even moments when the auxiliary variable contributes excessively to the error. This is unfavorable for the convergence of the solved variables and can also result in the actual convergence time exceeding the theoretically calculated value.
\par However, our current method still has space for improvement. First, we still need to estimate the two hyperparameters $\alpha$ and $\beta$ in advance. While these hyperparameters can be estimated for PDE solving problems, this process may not be feasible for general linear systems. Therefore, we need to develop algorithms with unified parameters rather than requiring predefined hyperparameters. Second, we use fixed hyperparameters, which creates a conflict between achieving rapid initial convergence and maintaining stability in later stages. How to design adaptive parameter algorithms for solving linear systems remains an open question that requires further discussion. Finally, the quantum algorithm we proposed is based on the assumption of sparse matrices. How to extend the algorithm to dense matrices is also a question worth considering  \cite{Wossnig2018Dense}.
%\addcontentsline{toc}{section}{Code Availability}
\section*{Code Availability}
\par The code that support the findings of the main text and the supplement information are will be publicly available upon acceptance.
%The code that support the findings of the main text and the supplement information are openly available in GitHub, \url{https://github.com/XiaoyangHe1997/Improved-Quantum-IMEX/tree/master}.

%\addcontentsline{toc}{section}{Declaration of competing interest}
\section*{Declaration of competing interest}
The authors declare that they have no known competing financial interests or personal relationships that could have appeared to influence the work reported in this paper.

%\addcontentsline{toc}{section}{Acknowledgement}
\section*{Acknowledgement}

SJ was supported by NSFC grants No. 12341104 and 92270001, Shanghai Science and Technology Innovation Action Plan 24LZ1401200,
the Shanghai Jiao Tong University 2030 Initiative, and the Fundamental Research
Funds for the Central Universities. 
XDZ was partly supported by the National Natural Science Foundation of China (No.12371354) and the Science and Technology Commission
of Shanghai Municipality, China (No.22JC1403600) and the Montenegrin Chinese Science and Technology (No.4-3).

\addcontentsline{toc}{section}{References}
%\bibliographystyle{alpha}
%\bibliography{reference}
\newcommand{\etalchar}[1]{$^{#1}$}

\end{document}